# Nanophotonic Cooling of Free Electrons Without Active Feedback

**Shiran Even-Haim[†1], Dor Gelem[†1], Aviv Karnieli[†1], Renwen Yu[3], Ido Kaminer[1], Vladan Vuletić[2], Shai Tsesses[*2]**

[1]*Faculty of Electrical & Computer Engineering and The Helen Diller Quantum Center, Technion - Israel Institute of Technology, Haifa 3200003, Israel*

[2]*Department of Physics and Research Laboratory of Electronics, Massachusetts Institute of Technology, Cambridge, Massachusetts 02139, USA*

[3]*Department of Electrical Engineering, Ginzton Laboratory, Stanford University, Stanford, California 94305, USA*

[†]*equal contributors*

*tsesses@mit.edu

**Free-electron beams are vital to microscopy, spectroscopy and material science, yet their utility remains limited by their intrinsic energy spread. This uncertainty limits electron spectroscopy and creates chromatic aberration in microscopy. We propose a method to reduce this energy uncertainty using frequency-dependent spontaneous photon emission, controlled by a dispersion-engineered photonic structure. Our method is active-feedback-free; resembles cavity-assisted laser cooling of atoms; and outperforms existing electron monochromators in ideal conditions. Our findings could improve electron microscopy and spectroscopy, potentially paving the way for precision metrology with electron beams.**

## 1. Introduction

Electron microscopy and spectroscopy have become indispensable tools for probing matter at the atomic scale, offering superior spatial resolution [1] combined with energy sensitivity [2]. Unfortunately, the performance of these important techniques is unavoidably constrained by the intrinsic energy uncertainty of available electron sources [3]. This energy spread limits the resolution of electron energy-loss spectrometers, such as those operating within transmission electron microscopes (TEMs), masking subtle excitonic [4] or phononic [5] signatures in materials. The scale of this spread is related to the mechanism for electron emission, and is typically on the order of the material work function: $\Delta E = 0.2 - 1$ eV [6,7].

Chromatic aberration is another consequence of this electron energy uncertainty, and since it cannot currently be corrected simultaneously with geometric aberrations, the chromatic aberration limits the ultimate spatial resolution of electron microscopes [8–10]. This issue is further exacerbated in microscopes using lower-energy electrons, such as scanning electron microscopes (SEMs), as the relative energy uncertainty increases [11].

The conventional strategy to overcome this limitation is to employ electro- and magneto-static monochromators [12,13] or energy filters [14]. These devices spectrally select a narrow portion of the electron energy distribution, down to sub-0.01 eV [15], yet contain a fundamental tradeoff between energy spread and electron flux [6,12,16]. Typically, monochromators transmit only 1–10% of the original beam current [17], which defines the extent to which they can reduce the electron energy spread. Aside from its high cost, this strategy is impractical for many microscopes due to its sensitivity to electromagnetic or mechanical instabilities [13]. Therefore, monochromators are very scarce for any electron beam that is outside of TEM energy scales.

Considering the importance of this problem, various other approaches to reduce the energy spread $\Delta E$ have been explored. One path to achieve this goal is partially reversing energy chirps introduced by the electron source via RF [18,19], THz [20], or optical [21,22] compression cavities, yet their performance thus far has not exceeded that of monochromators. Other recent methods are able to utilize ultrafast lasers and integrated photonic circuits to improve energy resolution [23–25] or chromatic aberration [26] beyond the state-of-the-art, albeit without reducing $\Delta E$, limiting them only to certain applications. There thus remains a critical need for a universal monochromation strategy that directly reduces the energy spread, for continuous and pulsed beams alike, across the large range of electron energies used in microscopy and spectroscopy.

The ideal solution to this problem would be to systematically narrow down the energy spread of the electron ensemble via a lossless process, i.e. to cool the electrons. Unlike filtering, cooling shrinks the phase-space volume of the ensemble – a reduction forbidden by Liouville's theorem for any non-dissipative dynamics – and therefore requires an energy-selective process that reduces the electron entropy. Cooling beams of electrons is a long-standing goal, motivated primarily by the need to improve electron beam brightness, and has mostly relied on cooling the electron source [27,28] or cooling the beam through radiation emission, usually over repeated interaction cycles [29–33] and typically with ultra-relativistic beams. Cooling by radiation emission has also been immensely successful in atoms [34,35], where the motional uncertainty of atoms has been reduced by addressing their internal energy structure with narrow-linewidth lasers.

Inspired by this paradigm, and the recent introduction of two-level system behavior in the interaction of free electrons with guided electromagnetic waves [36,37], we propose an analogous mechanism to cavity-assisted laser cooling [38–40] for free electrons based on their *spontaneous* emission into a photonic structure [41]. The phase-matching condition between the electrons and the mode of the photonic structure is designed such that higher (lower) energy electrons will spontaneously emit higher (lower) energy photons, as compared to the photons emitted by electrons with the mean beam energy, thus compressing the electron energy distribution after emission. We derive analytic expressions for the cooling mechanism, numerically explore its performance, and discuss its similarity to atomic cavity cooling.

In principle, the proposed radiative cooling can be performed with minimal loss of electrons, and with a lower limit on the energy spread that is orders-of-magnitude below the current state-of-the-art. Practically, we analyze the limitations on the efficiency and implementation of our scheme, while outlining the parameter regimes required to deploy it across various systems, ranging from low (~100 eV) to high (~GeV) energies. Our results are ubiquitous to any charged elementary particle (including, e.g., muons), are broadly applicable to electron microscopy or spectroscopy, and – given a sufficient reduction of electron energy uncertainty – would begin to allow metrology with electron beams, the likes of which has only been possible with trapped electrons [42–44], ions [45–47] or atoms [48,49].

## 2. Nanophotonic Free-Electron Cooling

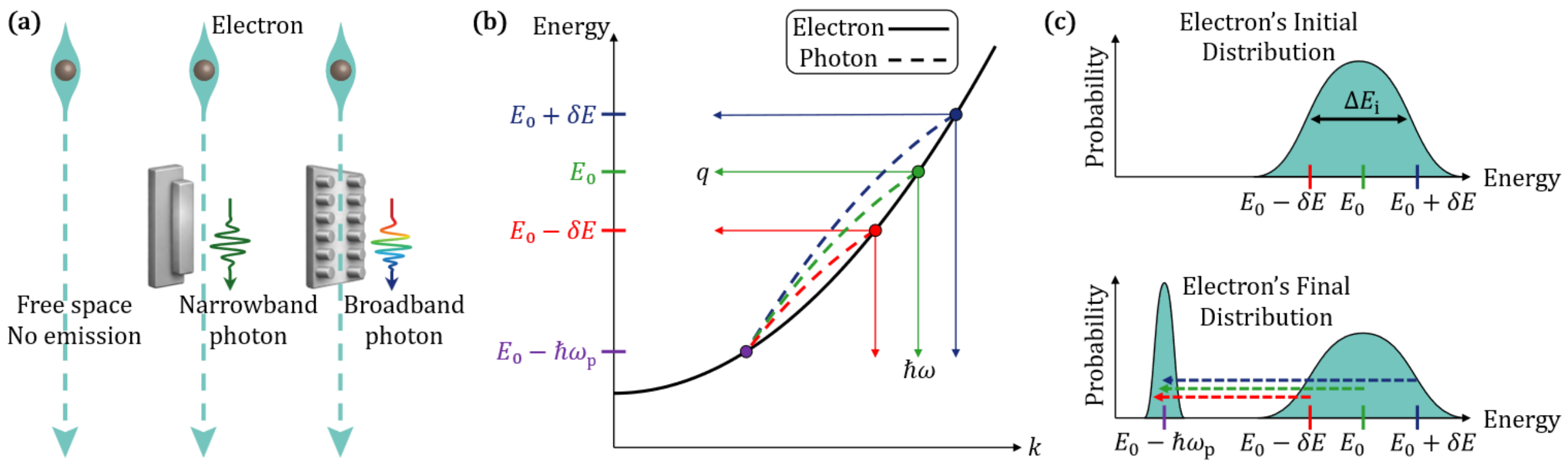

**Fig. 1: The Nanophotonic Free-Electron Cooling Mechanism. (a)** A free electron in three configurations. In free space (left), no photon is emitted and the energy distribution is unchanged. Near a structure providing phase-matching alone (middle), every emitting electron loses the same photon energy, so the distribution is shifted but not narrowed. Near an appropriately dispersion-engineered structure (right, this work), the frequency of the spontaneously emitted photon depends on the electron energy; the photon is broadband and carries away the electron's entropy. **(b)** Electron dispersion $E(k)$, with circles marking the initial states $E_0 - \delta E$, $E_0$ and $E_0 + \delta E$ (red, green, blue) and the final state $E_0 - \hbar\omega_p$ (purple). Dashed curves: the dispersion $\omega(q)$ of the photonic mode, drawn from each initial state on the colored axes (photon momentum leftward, photon energy downward). Emission is phase-matched where a dashed curve meets the electron dispersion, i.e., where the electron momentum loss $k(E) - k(E - \hbar\omega)$ equals the photon momentum $q(\omega)$. The dispersion was engineered such that all initial electron states meet the electron dispersion at the same point, so that the final electron energy after emission is always $E_0 - \hbar\omega_p$. **(c)** The final electron distribution consists of two parts: electrons that did not emit remain unchanged and centered around energy $E_0$, whereas the electrons that did emit are cooled and centered around energy $E_0 - \hbar\omega_p$. Throughout the manuscript, we assume that emission beyond a single event is unlikely (perturbative regime).

Our cooling mechanism relies on electron interaction with a photonic structure, causing the spontaneous emission of a photon into a guided mode, as illustrated in Fig. 1a. If the electron moves at a constant velocity in free space, photon emission is forbidden due to momentum conservation. However, a photonic structure can mitigate the momentum gap between a free electron and a photon, causing the electron to spontaneously emit when its group velocity matches the phase velocity of photonic modes in the structure [50–52]. When emitting to guided photonic modes, the electron predominantly emits in the direction of its propagation [52]. The microscopic mechanism for this process is interaction between the electron and the dipoles residing in the structure, creating a polarization field inside [53,54]. However, phase-matching alone causes electrons with different initial energies to end up in different final energies after photon emission, at most preserving the original electron energy distribution. Conversely, our scheme utilizes dispersion engineering together with phase-matching, such that the final energy of all emitting electrons will be the same.

The electron energy density after emitting a single photon, referred to as the "cooled" portion of the electron energy distribution, is shifted by $\hbar\omega$, denoting $\omega$ as the emitted photon frequency ($\hbar$ is the reduced Planck's constant). We analyze the case where the electron's initial energy density is normally distributed around $E_0$, with an energy spread at half-maximum (FWHM) given by $\Delta E_{\mathrm{i}}$, such that $\Delta E_{\mathrm{i}} \ll E_0$. Therefore, the mean energy of the cooled electron after interaction becomes $E_0 - \hbar\omega_{\mathrm{p}}$, where $\omega_{\mathrm{p}}$ is the mean emitted photon frequency, satisfying $\hbar\omega_{\mathrm{p}} > \Delta E_{\mathrm{i}}$ to ensure that the cooled portion of the electron energy distribution lies outside the initial energy spread. These conditions are illustrated in Figs. 1b,c.

For simplicity, we consider here only a perturbative electron-photon interaction where the normalized electron-light coupling constant satisfies $|g_{\mathrm{Q}}| \ll 1$, such that more than one emission event is unlikely. Additionally, the photonic structure is assumed to allow a single transverse mode family to dominantly interact with the electron around the phase matching point at $\omega_{\mathrm{p}}$. This is readily achieved via dispersion or spatial overlap engineering [23,37,51,52,55–59]. The cooling process is driven purely by spontaneous emission of the electron into the photonic mode at vacuum, i.e., no energy is inserted into the system and no active feedback is applied. A vacuum state of zero photons in the photonic mode is readily achieved, since for the frequencies $\omega_{\mathrm{p}}$ considered throughout this manuscript, thermal occupation is exponentially suppressed at room temperature.

Following the established formalism for free-electron–light interaction [57,60–65], the first-order transition matrix element from initial electron energy $E_{\mathrm{i}}$ to final electron energy $E_{\mathrm{f}}$ via spontaneous emission of a photon with frequency $\omega$ is (see SM S1.1-S1.4):

$$M_{E_{\mathrm{i}}\rightarrow E_{\mathrm{f}},\omega} = -\frac{i}{\hbar}\int_{-\infty}^{\infty} dt\, e^{-i\frac{E_{\mathrm{i}}-E_{\mathrm{f}}-\hbar\omega}{\hbar}t}\langle E_{\mathrm{f}}, 1_{\omega}|H_{\mathrm{int}}|E_{\mathrm{I}}, 0_{\omega}\rangle, \quad (1)$$

where $H_{\mathrm{int}} = \left(\frac{e}{m}\right)\mathbf{A}\cdot\mathbf{p}$ is the electron-photon interaction Hamiltonian, $e$ is the electron charge, $m$ is the electron mass, $\mathbf{A}$ is the vector potential of the photonic mode and $\mathbf{p}$ is the electron momentum operator. We can now define the phase mismatch: $\Delta k = k(E_{\mathrm{i}}) - k(E_{\mathrm{f}}) - q(\omega)$, where $\hbar q(\omega)$ is the momentum of the photon and $\hbar k(E)$ is the electron's momentum at energy $E$. The electron propagates adjacent to the photonic structure for an effective distance $L_{\mathrm{int}}$. Therefore, Eq. (1) can be rewritten as a product of a Dirac delta distribution in energy, enforcing energy conservation, and a sinc function $\left(\mathrm{sinc}(x) = \frac{\sin(x)}{x}\right)$ distribution in the phase mismatch $\Delta k$, relaxing momentum conservation (see SM S1.4):

$$M_{E_{\mathrm{i}}\rightarrow E_{\mathrm{f}},\omega} = -i\sqrt{\frac{k(E_{\mathrm{i}})}{k(E_{\mathrm{f}})}}\sqrt{\frac{e^2}{4\pi\epsilon_0\hbar\omega v_{\mathrm{p,g}}}}\, g_{\mathrm{eff},\omega}\,\delta(E_{\mathrm{i}} - E_{\mathrm{f}} - \hbar\omega)L_{\mathrm{int}}\mathrm{sinc}\left(\Delta k\frac{L_{\mathrm{int}}}{2}\right). \quad (2)$$

Here, $\epsilon_0$ is the vacuum permittivity, $g_{\mathrm{eff},\omega} = \int d^2\boldsymbol{r}_{\perp}|\psi_{\perp}(\boldsymbol{r}_{\perp})|^2\tilde{u}_{\omega}^{(z)*}(\boldsymbol{r}_{\perp})$ is the effective spatial overlap coefficient between the normalized transverse electron wavefunction $\psi_{\perp}$ and the normalized transverse profile $\tilde{u}_{\omega}^{(z)}$ of the longitudinal field of the photonic mode (SM S1.4), and the photonic group velocity is $v_{\mathrm{p,g}} = \left(\left.\frac{\partial q}{\partial\omega}\right|_{\omega_{\mathrm{p}}}\right)^{-1}$. The energy conservation in Eq. (2) is a matter of convenience, as it could also be written with momentum conservation and energy uncertainty. In practice, for a finite system, there are both energy and momentum uncertainties, as shown in Fig. 2a. The final electron energy distribution density matrix $\rho_{\mathrm{f}}(E_{\mathrm{f}})$ can now be written in terms of the initial electron density matrix $\rho_{\mathrm{i}}(E_{\mathrm{I}}, E_{\mathrm{i}}')$ and the transition matrix element from Eq. (2) (SM S1.5):

$$\rho_{\mathrm{f}}(E_{\mathrm{f}}) = \frac{e^2L_{\mathrm{int}}^2}{4\pi\epsilon_0\hbar v_{\mathrm{p,g}}}\int d\omega\,\frac{k(E_{\mathrm{f}}+\hbar\omega)}{\omega k(E_{\mathrm{f}})}\left|g_{\mathrm{eff},\omega}\right|^2\mathrm{sinc}^2\left(\Delta k\frac{L_{\mathrm{int}}}{2}\right)\rho_{\mathrm{i}}(E_{\mathrm{f}}+\hbar\omega, E_{\mathrm{f}}+\hbar\omega). \quad (3)$$

Our cooling mechanism relies on the interaction length limiting momentum conservation violation, represented by the phase mismatch $\Delta k$, and Eq. (3) shows that meaningful contributions to the transition occur when $|\Delta k|L_{\mathrm{int}} \lesssim 2\pi$. The $\mathrm{sinc}^2\left(\Delta k\frac{L_{\mathrm{int}}}{2}\right)$ term selects a phase-matching window of final energies around the zero of $\Delta k$, with FWHM of $\Delta E_{\mathrm{f}}^{\mathrm{win}} = \frac{2\mu}{L_{\mathrm{int}}\left|\frac{d\Delta k}{dE_{\mathrm{f}}}\right|_{\Delta k=0}}$, where $\mu = 2.783$ is the

FWHM of $\mathrm{sinc}^2 x$ in its dimensionless argument, and the derivative is evaluated for strict momentum conservation, $\Delta k = 0$. All numerical results in Fig. 2b,c & 3 below are calculated from Eq. (3), and therefore include the full electron dispersion while expanding the photonic dispersion to second order around $\omega_\mathrm{p}$ within a fixed photonic bandwidth (SM S4.5).

We evaluate $\left|\frac{d\Delta k}{dE_\mathrm{f}}\right|_{\Delta k=0}$ analytically by expanding $\Delta k$ around the phase-matching point, $k(E_0) = k(E_0 - \hbar\omega_\mathrm{p}) + q(\omega_\mathrm{p})$, and defining deviations of the electron initial and final energy, as well as the emitted photon frequency: $\delta E_\mathrm{i} = E_\mathrm{i} - E_0$, $\delta E_\mathrm{f} = E_\mathrm{f} - E_0 + \hbar\omega_\mathrm{p}$, $\delta\omega = \omega - \omega_\mathrm{p}$, such that energy conservation enforces $\delta E_\mathrm{i} = \delta E_\mathrm{f} + \hbar\delta\omega$. It is then instructive to expand the phase mismatch $\Delta k$ to second order in $\delta\omega$ and first order in $\delta E_\mathrm{f}$ (SM S2.1):

$$\Delta k = \left[\frac{\Delta v_\mathrm{i-f}^{-1}}{\hbar} + \left(\left.\frac{\partial^2 k}{\partial E^2}\right|_{E_0}\right)\hbar\delta\omega\right]\delta E_\mathrm{f} + \Delta v_\mathrm{GVM}^{-1}\delta\omega + \frac{1}{2}\left[\left.\frac{\partial^2 k}{\partial E^2}\right|_{E_0} - \frac{1}{\hbar^2}\left(\left.\frac{\partial^2 q}{\partial \omega^2}\right|_{\omega_\mathrm{p}}\right)\right]\hbar^2\delta\omega^2 + O(\delta E_\mathrm{f}^2) + O(\delta\omega^3). \tag{4}$$

Here, $\Delta v_\mathrm{GVM}^{-1} = v_\mathrm{e}^{-1} - v_\mathrm{p,g}^{-1}$ is the reciprocal group velocity mismatch between the electron and the photon, and $\Delta v_\mathrm{i-f}^{-1} = v_\mathrm{e}^{-1} - v_\mathrm{e,f}^{-1}$ is the reciprocal group velocity difference of the electron before and after emission, with $v_\mathrm{e} = \left(\hbar\left.\frac{\partial k}{\partial E}\right|_{E_0}\right)^{-1}$ and $v_\mathrm{e,f} = \left(\hbar\left.\frac{\partial k}{\partial E}\right|_{E_0-\hbar\omega_\mathrm{p}}\right)^{-1}$. Terms of order $\delta E_\mathrm{f}^2$ are neglected, as they contribute a relative correction of at most $\sim \delta E_\mathrm{f}/(E_0 - \hbar\omega_\mathrm{p})$ to the linear term (SM S2.2), which is small since $|\delta E_\mathrm{f}| \lesssim \Delta E_\mathrm{i} \ll E_0 - \hbar\omega_\mathrm{p}$.

Optimal cooling is achieved when momentum conservation ($\Delta k = 0$) forces $\delta E_\mathrm{f} \to 0$ for every emitted photon frequency $\omega_p + \delta\omega$, reducing the electron to a single final energy, $E_0 - \hbar\omega_\mathrm{p}$. Two conditions are thus imposed on the photonic mode at $\omega_\mathrm{p}$ in addition to phase matching: First, the photon group velocity must match the electron velocity, $v_\mathrm{p,g} = v_\mathrm{e}$ (i.e., $\Delta v_\mathrm{GVM}^{-1} = 0$), eliminating the term linear in $\delta\omega$. Second, the photonic group-velocity dispersion must be engineered to cancel the photon recoil applied on the electron, such that $\left.\frac{\partial^2 q}{\partial\omega^2}\right|_{\omega_\mathrm{p}} = \hbar^2\left.\frac{\partial^2 k}{\partial E^2}\right|_{E_0}$. Each electron therefore emits precisely the photon that carries away its own energy deviation, $\hbar\delta\omega = \delta E_\mathrm{i}$, with the initial energy spread transferred to the photon spectrum and the emitting electrons collapsing onto $E_0 - \hbar\omega_\mathrm{p}$, up to an uncertainty related to the finite interaction length and high-order terms in $\delta\omega$ (see Fig. 2a).

When the group velocities are not matched, momentum conservation ($\Delta k = 0$) correlates $\delta E_\mathrm{f}$ with the emitted frequency:

$$\left[\Delta v_{\text{i-f}}^{-1} + \left(\frac{\partial^2 k}{\partial E^2}\bigg|_{E_0}\right)\hbar^2\delta\omega\right]\delta E_{\text{f}} + \Delta v_{\text{GVM}}^{-1}\hbar\delta\omega = 0, \tag{5}$$

with $\delta\omega^2$ terms canceled through dispersion engineering, and higher-order terms being negligible. Combining Eq. (5) with energy conservation gives, up to first order, the following relation between the initial and final energy distribution:

$$\frac{\delta E_{\text{f}}}{\delta E_{\text{i}}} \approx \frac{\Delta v_{\text{GVM}}^{-1}}{\Delta v_{\text{GVM}}^{-1} - \Delta v_{\text{i-f}}^{-1}} = \frac{v_{\text{e}}^{-1} - v_{\text{p,g}}^{-1}}{v_{\text{e,f}}^{-1} - v_{\text{p,g}}^{-1}}. \tag{6}$$

Eq. (6) shows that optimal cooling is achieved by group-velocity matching, $v_{\text{e}} = v_{\text{p,g}}$, nullifying the dependence of the final electron energy on the initial energy (this is corroborated numerically in Fig. 2b). The general condition for cooling is $\left|\frac{\delta E_{\text{f}}}{\delta E_{\text{i}}}\right| < 1$, resulting in the condition:

$$v_{\text{e}}^{-1} + v_{\text{e,f}}^{-1} > 2v_{\text{p,g}}^{-1}. \tag{7}$$

For a non-relativistic electron: $k(E) = \sqrt{\frac{2m}{\hbar^2}E}$, and Eq. (7) can be expressed directly in terms of the velocities and photon wavelength. Energy conservation at the mean emission frequency gives a final electron group velocity $v_{\text{e,f}} = \sqrt{v_{\text{e}}^2 - \frac{4\pi\hbar c}{m\lambda_{\text{p}}}}$, where $\lambda_{\text{p}} = 2\pi c/\omega_{\text{p}}$. For $\frac{4\pi\hbar c}{m\lambda_{\text{p}}} < v_{\text{p,g}}^2$, which allows group-velocity matching, the cooling condition of Eq. (7) becomes $v_{\text{e}} < v_{\text{p,g}}\left(1 + \frac{\pi\hbar c}{m\lambda_{\text{p}}v_{\text{p,g}}^2} + \frac{\pi^2\hbar^2c^2}{m^2\lambda_{\text{p}}^2v_{\text{p,g}}^4} + \cdots\right)$. Photon recoil therefore shifts the cooling–heating boundary slightly above $v_{\text{e}} = v_{\text{p,g}}$, while leaving $v_{\text{e}} = v_{\text{p,g}}$ as the optimal cooling point, as shown in Fig. 2b. The denominator of Eq. (6) vanishes when $v_{\text{e,f}} = v_{\text{p,g}}$, producing a heating resonance on the $v_{\text{e}} > v_{\text{p,g}}$ side (also corroborated in Fig. 2b). The finite phase-matching bandwidth, the finite photonic bandwidth and higher-order terms in Eq. (4) regularize this apparent divergence (SM S2.3). Fig. 2c shows that decreasing $\lambda_{\text{p}}$, i.e., increasing the photon recoil, both increases the cooling factor and shifts the phase-matching condition, similarly to its role in atomic laser cooling [34,35].

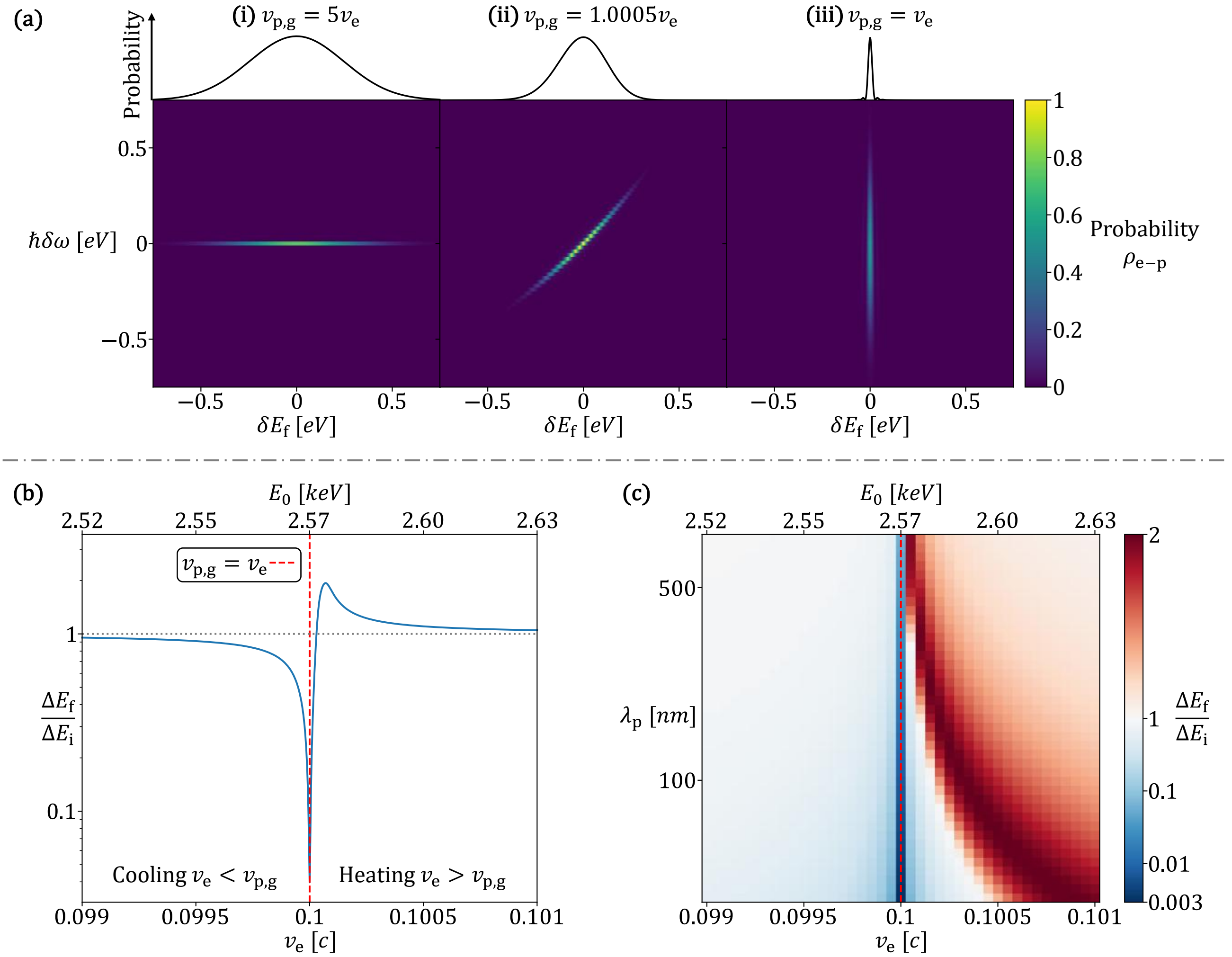

**Fig. 2: Velocity and wavelength dependence of nanophotonic free-electron cooling. (a)** Three electron velocity conditions are compared: **(i)** phase matching only ($v_{\mathrm{p,g}} = 5v_{\mathrm{e}}$), **(ii)** partial group-velocity matching ($v_{\mathrm{p,g}} = 1.0005v_{\mathrm{e}}$) and **(iii)** full dispersion engineering ($v_{\mathrm{p,g}} = v_{\mathrm{e}}$). the figure shows the electron energy probability (top) and the joint electron-photon probability density $\rho_{\mathrm{e-p}}$ (bottom) for an emitting electron as a function of the final electron energy deviation $\delta E_{\mathrm{f}}$ and the photon energy deviation $\hbar\delta\omega$, calculated via Eq. (3). In option (i), the electron keeps its initial width; in option (ii), the probability density rotates and the electron energy spread is twice as low; in option (iii), the final electron energy spread is ~25 times as low. In (i) and (ii) the curves are multiplied by 24 and 12, respectively, so that all curves will have the same height. Parameters used are $\Delta E_{\mathrm{i}} = 0.584\ eV$; $v_{\mathrm{e}} = 0.1c$; $E_0 = 2.57\ keV$; $\lambda_{\mathrm{p}} = 500\ nm$; and $L_{\mathrm{int}} = 0.01\ m = 2 \cdot 10^4 \lambda_{\mathrm{p}}$. **(b)** Electron cooling ratio $\frac{\Delta E_{\mathrm{f}}}{\Delta E_{\mathrm{i}}}$ as a function of the mean electron velocity $v_{\mathrm{e}}$ (top axis: the corresponding mean energy $E_0$), with the mode phase-matched at $\lambda_{\mathrm{p}}$ for every $v_{\mathrm{e}}$, where $\Delta E_{\mathrm{f}}$ is the FWHM of the final electron energy distribution calculated from Eq. (3). A final FWHM of $\Delta E_{\mathrm{f}} = 0.023\ eV$ ($\frac{\Delta E_{\mathrm{f}}}{\Delta E_{\mathrm{i}}} = 0.04$) is calculated using the parameters above at $v_{\mathrm{e}} = v_{\mathrm{p,g}}$. The asymmetric (Fano-like) line shape is analogous to the detuning dependence of atomic laser cooling. The red dashed line marks the optimal group-velocity-matching point $v_{\mathrm{e}} = v_{\mathrm{p,g}}$. Photon recoil shifts the cooling–heating boundary slightly to $v_{\mathrm{e}} > v_{\mathrm{p,g}}$, while the heating resonance occurs near $v_{\mathrm{e,f}} = v_{\mathrm{p,g}}$. **(c)** Electron cooling ratio $\frac{\Delta E_{\mathrm{f}}}{\Delta E_{\mathrm{i}}}$ as a function of $v_{\mathrm{e}}$ and the phase-matched photon wavelength $\lambda_{\mathrm{p}}$, for the same $v_{\mathrm{p,g}}$, $\Delta E_{\mathrm{i}}$ and $L_{\mathrm{int}}$. The red dashed line marks the optimal group-velocity-matching point $v_{\mathrm{e}} = v_{\mathrm{p,g}}$.

At the group-velocity-matching point $v_{\mathrm{e}} = v_{\mathrm{p,g}}$, the first-order mapping contribution in Eq. (6) vanishes, and the final FWHM is given solely by the phase-matching window. Neglecting higher-order terms, the cooling ratio is:

$$\frac{\Delta E_{\mathrm{f}}}{\Delta E_{\mathrm{i}}} \simeq \frac{2\mu\hbar}{L_{\mathrm{int}}\left|\Delta v_{\mathrm{i-f}}^{-1}\right|\Delta E_{\mathrm{i}}} = \frac{L_0}{L_{\mathrm{int}}}, \qquad L_0 \equiv \frac{2\mu\hbar}{\left|\Delta v_{\mathrm{i-f}}^{-1}\right|\Delta E_{\mathrm{i}}}. \tag{8a}$$

The emitted electrons are therefore narrower than the initial distribution only above a threshold interaction length $L_0$ (see Fig. 3a). Cooling at matched velocity therefore requires:

$$L_{\mathrm{int}} > L_0 = \frac{2\mu\hbar}{\left|\Delta v_{\mathrm{i-f}}^{-1}\right|\Delta E_{\mathrm{i}}}. \tag{8b}$$

For a non-relativistic electron, $L_0 = \frac{2\mu\hbar}{\left[\left(v_{\mathrm{e}}^2 - \frac{4\pi\hbar c}{m\lambda_{\mathrm{p}}}\right)^{-1/2} - v_{\mathrm{e}}^{-1}\right]\Delta E_{\mathrm{i}}}$ (SM S2.2); the same analysis applies to relativistic electrons, for which only the electron dispersion changes (SM S5). The final energy spread thus increases with the photon wavelength and with the electron energy, favoring low-energy electrons and short photon wavelengths. Away from group-velocity matching point ($v_{\mathrm{e}} \neq v_{\mathrm{p,g}}$), the cooling ratio approaches the $L_{\mathrm{int}}$-independent value of Eq. (6) at long interaction lengths. The heatmap in Fig. 3b demonstrates this numerically, and its $\Delta E_{\mathrm{f}} = \Delta E_{\mathrm{i}}$ contour gives the numerical cooling boundary, coinciding with the analytical expression for $L_0$.

The results of Figs. 2,3 suggest that our method can, in principle, cool electrons to any desirable energy spread, but unfortunately, photonic propagation loss broadens the phase-matching response inherently and therefore imposes a lower limit on the attainable energy spread. Treating the lossy photonic structure within macroscopic quantum electrodynamics [66,67] (SM S3 and S4.1), the final electron energy distribution of Eq. (3) acquires the Lorentzian line shape of the lossy photonic mode:

$$\rho_{\mathrm{f}}(E_{\mathrm{f}}) = \frac{e^2 L_{\mathrm{int}}^2}{4\pi\epsilon_0\hbar v_{\mathrm{p,g}}}\int d\omega\,\frac{k(E_{\mathrm{f}}+\hbar\omega)}{k(E_{\mathrm{f}})}\rho_{\mathrm{i}}(E_{\mathrm{f}}+\hbar\omega, E_{\mathrm{f}}+\hbar\omega)$$
$$\times \int d\omega' \left|g_{\mathrm{eff},\omega'}\right|^2 \mathrm{sinc}^2\left[\Delta k(E_{\mathrm{f}},\omega')\frac{L_{\mathrm{int}}}{2}\right]\frac{1}{\pi\omega'}\frac{\alpha/2}{(\omega'-\omega)^2+(\alpha/2)^2}. \tag{9}$$

Here, $\alpha$ represents the loss rate of the photonic mode. It is related to the propagation loss in dB/cm used in Fig. 3 through the longitudinal attenuation factor $e^{-\alpha z/v_{\mathrm{p,g}}}$. The effective interaction length is thus limited to $L_{\mathrm{loss}} = \int_0^{L_{\mathrm{int}}} dz e^{-\alpha z/v_{\mathrm{p,g}}} = \frac{v_{\mathrm{p,g}}}{\alpha}\left(1 - e^{-\alpha L_{\mathrm{int}}/v_{\mathrm{p,g}}}\right)$, such that the phase-matching bandwidth, and with it the final energy spread, can no longer be reduced by increasing $L_{\mathrm{int}}$ beyond the attenuation length $v_{\mathrm{p,g}}/\alpha$. At interaction lengths much longer than the attenuation length, the final energy spread saturates at $\Delta E_{\mathrm{f,sat}}$ (SM S4.2), which is independent of both $\Delta E_{\mathrm{i}}$ and $L_{\mathrm{int}}$, and at group-velocity matching is given by $\frac{\Delta E_{\mathrm{f,sat}}}{E_0} \simeq \frac{\alpha\lambda_{\mathrm{p}}}{\pi c}$. The loss-included $\Delta E_{\mathrm{f}}$ shown in Fig. 3a is calculated from Eq. (9) at group-velocity matching, $v_{\mathrm{e}} = v_{\mathrm{p,g}}$, and includes both the finite phase-matching bandwidth and the initial electron distribution. For the parameters of Fig. 3a, the floors are around 4.75, 14.2 and 47.5 meV for propagation losses of 5, 15 and 50 dB/cm, respectively. for relativistic electrons, $\Delta E_{\mathrm{f,sat}}$ increases by $\gamma(\gamma+1)/2$.

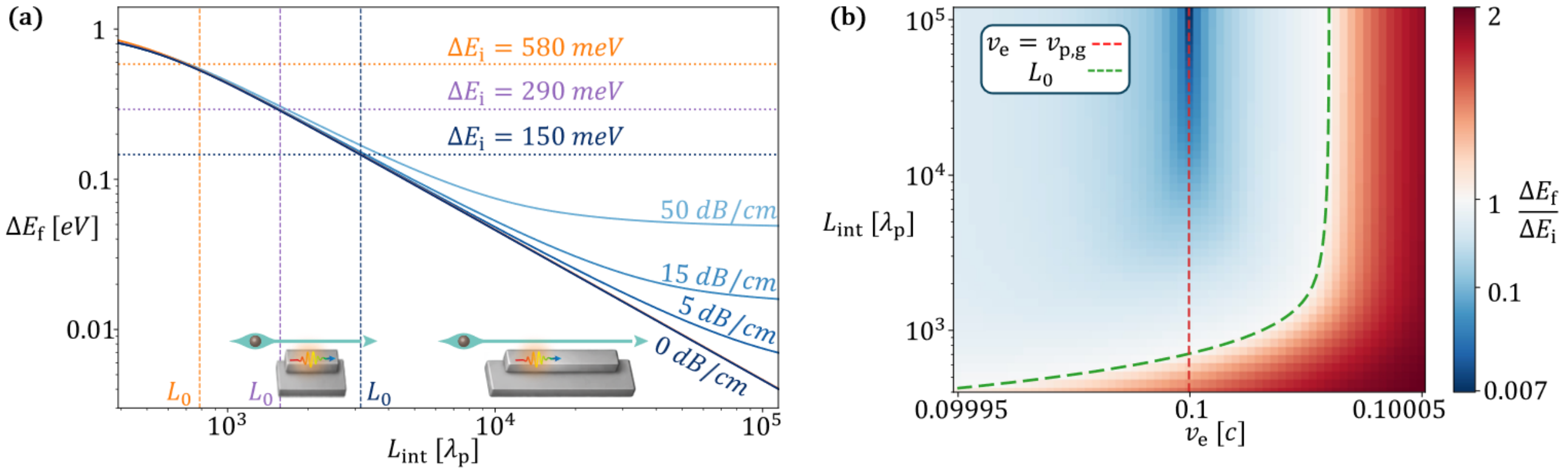

**Fig. 3: Finite interaction-length and photonic-loss limits. (a)** Final electron energy spread $\Delta E_{\mathrm{f}}$ (FWHM) as a function of $L_{\mathrm{int}}/\lambda_{\mathrm{p}}$ at $v_{\mathrm{e}} = v_{\mathrm{p,g}} = 0.1c$ and $\lambda_{\mathrm{p}} = 500\,\mathrm{nm}$. The lossless curves are shown for $\Delta E_{\mathrm{i}} = (0.584, 0.292, 0.146)$ eV. The vertical dashed lines mark the corresponding cooling threshold lengths $L_0$ from Eq. $(8b)$, $L_0 = (786{,}1573{,}3145)\cdot\lambda_{\mathrm{p}}$, and the horizontal dotted lines mark the initial energy spreads. Beyond $L_0$, all lossless curves collapse onto the $\Delta E_{\mathrm{f}} \propto 1/L_{\mathrm{int}}$ dependence of Eq. $(8a)$. At shorter lengths, the final spread is limited by the assumed photonic bandwidth of 1 eV (FWHM, raised-cosine edges). For $\Delta E_{\mathrm{i}} = 0.146\,\mathrm{eV}$, the curves labeled 5, 15, and 50 dB/cm show the saturation $\Delta E_{\mathrm{f,sat}}$ caused by photonic propagation loss. **(b)** Cooling ratio $\Delta E_{\mathrm{f}}/\Delta E_{\mathrm{i}}$ as a function of $v_{\mathrm{e}}$ and $L_{\mathrm{int}}/\lambda_{\mathrm{p}}$, calculated from Eq. (3) without propagation loss, for $v_{\mathrm{p,g}} = 0.1c$, $\lambda_{\mathrm{p}} = 500$ nm and $\Delta E_{\mathrm{i}} = 0.584$ eV, with the same 1 eV photonic bandwidth. The red dashed line marks $v_{\mathrm{e}} = v_{\mathrm{p,g}}$, the green dashed line marks the threshold length $L_0(v_{\mathrm{e}})$, defined by $\Delta E_{\mathrm{f}} = \Delta E_{\mathrm{i}}$ (white region of the map).

Eq. (9) draws an interesting comparison between our method and cavity-assisted cooling of atoms [38–40]. The phase matching condition and bandwidth in our case are analogous to the atomic transition and cavity linewidth in the cavity cooling case, with electron velocity here playing a part similar to laser detuning. Both methods are limited by broadening originating from photonic losses, both have the same dependence on the photon recoil, and both are the result of a chromatically modified electromagnetic vacuum. There are, however, several marked differences: While cavity cooling changes the photonic density of states and is inherently tied to a passive feedback mechanism applied by the cavity, increasing the probability of emitting certain photon energies, here we change only the photonic dispersion. Intriguingly, one should be able to engineer the probability of emitting photons in our system as well, to reach a similar cooling effect (i.e., by engineering the electron-photon coupling $|g_{\mathrm{Q}}|$), but that notion is beyond the scope of our current manuscript.

# 3. Discussion and outlook

In summary, we presented a method that achieves an energy-selective, dissipative compression of the electron's energy distribution within a single-pass geometry. Our method relies on dispersion engineering of structural electromagnetic modes to match the electron group velocity in a wide bandwidth, and can potentially be amended to cool other useful charged particles, such as muons [68]. We showed that our scheme is, in principle, capable of monochromation beyond state-of-the-art techniques and in a comprehensive range of energy regimes, without depending on the selection of a small percentage of electrons. Furthermore, we discussed the analogy between the effect predicted herein and cavity-assisted laser cooling of atoms, with one major distinction: our method is active-feedback-free, unlike other recent methods to perform electron cooling [33,69], and does not require the added insertion of energy.

While our technique does not reduce energy uncertainty by lowering the number of electrons, it does have a finite efficiency, which is determined by the probability of a single photon emission. In this work, we assumed weak interaction for simplicity, but future work can extend the analysis to a non-perturbative regime [52,58], where emission of a single photon can be very efficient (10's of percent). This regime can further open a path to new cooling schemes based on multiple photon emission events by a single electron. Monochromation can also be further improved by concatenating several engineered structures with varying phase-matching points; utilizing repeated electron passage via an electron cavity or cycler; and dispersion engineering the nanophotonic structure to higher orders.

Table 1 in the End Matter estimates the cooling performance and the practical constraints in different electron energy regimes (see also SM S7). These constraints are technical challenges that may be overcome in the future. The fundamental limit of our method is tied to the lowest possible optical losses for the structure interacting with the electron beam. In this regard, fused silica fibers have already shown ultra-low optical losses, recently on the order of 0.14 dB/km [70], and special optical fibers have gone down to ~0.05 dB/km in certain wavelengths [71]. Assuming therefore a lower limit in the 0.1 dB/km range, the minimal electron beam energy spread possible with our method will be ~1 neV for a suitably chosen photon energy, irrespective of the electron energy and without compromising on the number of electrons.

Since its basic limitations are only related to the finite phase-matching bandwidth, bound by photonic propagation loss, our method has the potential to significantly improve chromatic aberration in electron microscopy and the resolution of electron spectroscopy. Beyond these "standard" applications, very low electron beam energy uncertainty would serve as an ideal platform for free-electron quantum optics [72], enabling the preparation of coherent states, entanglement, and single-electron interferometry with minimal decoherence. Additionally, since the electron transfers its initial energy uncertainty to the photon it emits, it can both produce and (via measurement) herald ultrafast single photon pulses. Most intriguingly, our method may open a route – however difficult – to reduce the energy uncertainty of electron beams below the metrologically-relevant level of ~$10^{-10}$, enabling table-top electron microscopes to probe fundamental constants in a relativistic regime.

During preparation of this manuscript we became aware of another, active-feedback-based method to perform free electron cooling with nanophotonic structures [69]. That said, our physical mechanisms are quite distinct, and so are our cooling limits, constraints and relevant photonic structures.

# End Matter

We examine several different electron energy regimes in Table 1, and consider practical achievable limits on the interaction lengths, limiting the cooling ratio (see SM S7). Our method can substantially improve both spatial and spectral resolution for low-energy electrons ($\sim 20 - 300\ eV$) and SEM-scale energies ($\sim 1 - 30\ keV$), by reducing energy spread and chromatic aberration, yet group-velocity-matching at these low energies requires slow-light modes that remain nearly constant over an ultrawide bandwidth. TEM-scale energies ($\sim 30 - 300\ keV$) are more favorable with respect to the requirements of the photonic modes but require far longer interaction lengths for a substantial improvement. In the ultra-relativistic accelerator regime ($\sim 1\ GeV$), energy spreads are much larger [73,74], keeping the distance required for a significant effect quite short. However, since cooling requires $\hbar\omega_{\mathrm{p}} > \Delta E_{\mathrm{i}}$ (Sec. 2), far higher photon energies must be used ($\gamma$-rays in this case).

| Regime | Electron energy $\boldsymbol{E_0}$ | Initial FWHM $\boldsymbol{\Delta E_{\mathrm{i}}}$ | Numerical final FWHM $\boldsymbol{\Delta E_{\mathrm{f}}}$ | Photon wavelength $\boldsymbol{\lambda_{\mathrm{p}}}$ | Interaction length $\boldsymbol{L_{\mathrm{int}}}$ | $\boldsymbol{L_{\mathrm{int}}/\lambda_{\mathrm{p}}}$ | Threshold cooling length $\boldsymbol{L_0}$ | Cooling factor $\boldsymbol{\Delta E_{\mathrm{i}}/\Delta E_{\mathrm{f}}}$ |
|---|---|---|---|---|---|---|---|---|
| Low Energy | 0.3 keV | 600 meV | 9.13 meV | 500 nm | 0.1 cm | $2 \cdot 10^3$ | 15 μm | 65.7 |
| SEM | 2.5 keV | 600 meV | 22.16 meV | 500 nm | 1 cm | $2 \cdot 10^4$ | 366 μm | 27.1 |
| TEM1 | 80 keV | 600 meV | 44.76 meV | 500 nm | 1 m | $2 \cdot 10^6$ | 7 cm | 13.4 |
| TEM2 | 200 keV | 150 meV | 41.01 meV | 500 nm | 5 m | $1 \cdot 10^7$ | 1.37 m | 3.66 |
| Ultra-relativistic | 1 GeV | 1 MeV | 33.42 keV | 0.1 pm | 1 cm | $1 \cdot 10^{11}$ | 330 μm | 29.9 |

**Table 1: Estimated cooling performance in different electron energy regimes,** assuming negligible losses of the photonic mode. Electron energies and initial FWHM energy spreads taken from reported experiments. The threshold length $L_0$ is given by Eq. $(8b)$ and the final FWHM energy spreads and cooling factor $\Delta E_{\mathrm{i}}/\Delta E_{\mathrm{f}}$ are numerically calculated from Eq. (3), assuming group-velocity matching $v_{\mathrm{e}} = v_{\mathrm{p,g}}$. The initial spread is determined by the type of electron source.

Each regime described above has its own significant challenge: broadband slow-light modes are difficult to engineer due to material properties; long interaction lengths are limited due to charging and beam divergences; and optical design in the $\gamma$-ray and x-ray range is nearly non-existent. That said, there are still several promising and practical platforms for a near-term demonstration of our predictions (see SM S7). For example, periodic fused silica or silicon nitride structures can potentially supply sub-meV energy spreads (~10 meV in more realistic scenarios) when designed for SEM energies and NIR photons, or for low energy electrons and UV photons. Current monochromators are difficult to operate at these energies, and our method is also superior in the amount of retained electrons at low energy uncertainty (see SM S6). Although the optical design is most favorable for optical photons and TEM energies, the optimal performance of our method in this regime, with practical parameters, does not exceed that of current monochromators.

# Nanophotonic Cooling of Free Electrons Without Active Feedback - Supplementary Material

**Shiran Even-Haim[†1], Dor Gelem[†1], Aviv Karnieli[†1], Renwen Yu[3], Ido Kaminer[1], Vladan Vuletić[2], Shai Tsesses[*2]**

[1]*Faculty of Electrical & Computer Engineering and The Helen Diller Quantum Center, Technion - Israel Institute of Technology, Haifa 3200003, Israel*

[2]*Department of Physics and Research Laboratory of Electronics, Massachusetts Institute of Technology, Cambridge, Massachusetts 02139, USA*

[3]*Department of Electrical Engineering, Ginzton Laboratory, Stanford University, Stanford, California 94305, USA*

[†]*equal contributors*

*tsesses@mit.edu

# Contents

# S1 Free Electron-Photon Interaction

## S1.1 Non-Relativistic Hamiltonian

We follow the established formalism for free-electron–light interaction [1–7] and summarize here only the ingredients and conventions used in this work. The total Hamiltonian governing a free electron interacting with the quantized electromagnetic field in a nanophotonic waveguide is:

$$H = H_{\mathrm{e}} \otimes \mathbb{I}_{\mathrm{ph}} + \mathbb{I}_{\mathrm{e}} \otimes H_{\mathrm{ph}} + H_{\mathrm{int}}, \tag{S1.1}$$

where $\otimes$ separates the electron and photon Hilbert spaces and $\mathbb{I}$ denotes the identity on the respective subspace. In the non-relativistic limit, the free-electron Hamiltonian is:

$$H_{\mathrm{e}} = \frac{\hat{\mathbf{p}}^2}{2m}, \tag{S1.2}$$

with $m$ the electron mass and $\hat{\mathbf{p}}$ the canonical momentum operator. The photon's Hamiltonian is:

$$H_{\mathrm{ph}} = \int_0^\infty d\omega\, \hbar\omega \hat{a}_\omega^\dagger \hat{a}_\omega\,, \tag{S1.3}$$

where $\hat{a}_\omega$ and $\hat{a}_\omega^\dagger$ are the annihilation and creation operators of the photonic mode at frequency $\omega$ (the zero-point energy is omitted), satisfying:

$$\left[\hat{a}_\omega, \hat{a}_{\omega'}^\dagger\right] = \delta(\omega - \omega'), \qquad [\hat{a}_\omega, \hat{a}_{\omega'}] = 0. \tag{S1.4}$$

Working in the Coulomb gauge ($\nabla \cdot \mathbf{A} = 0$), in which $\mathbf{A} \cdot \hat{\mathbf{p}} = \hat{\mathbf{p}} \cdot \mathbf{A}$, the interaction Hamiltonian obtained from minimal coupling is:

$$H_{\text{int}} = \frac{e}{m}\mathbf{A}(\mathbf{r}) \cdot \hat{\mathbf{p}}, \tag{S1.5}$$

where $e > 0$ is the elementary charge and $\mathbf{A}$ is the quantized vector potential. The $\mathbf{A}^2$ term is neglected: it is second order in $e$ and contributes only to two-photon processes, which lie beyond the first-order treatment used here.

The electron propagates along $z$ with transverse momentum small compared to the longitudinal momentum, $p_\perp \ll p_z$. In this paraxial regime $\hat{\mathbf{p}} \approx \hat{p}_z \hat{\mathbf{z}}$, so that $H_{\text{int}} \approx \frac{e}{m}\hat{A}_z(\mathbf{r})\hat{p}_z$, and the electron energy eigenstates take the factorized form:

$$\langle \mathbf{r}|E\rangle = \frac{\psi_\perp(\mathbf{r}_\perp)}{\sqrt{2\pi\hbar v(E)}} e^{ik(E)z}, \tag{S1.6}$$

with transverse normalization $\int d^2r_\perp |\psi_\perp(\mathbf{r}_\perp)|^2 = 1$ and longitudinal wavenumber:

$$k(E) = \frac{\sqrt{2mE}}{\hbar}, \qquad v(E) = \frac{\hbar k(E)}{m}. \tag{S1.7}$$

## S1.2 Photonic Mode Expansion and Normalization

As in Sec. 2 of the main text, we assume that the photonic structure allows a single transverse mode family to dominantly interact with the electron around the phase-matching point. The quantized vector potential of this photonic mode is written as a continuum of frequency modes:

$$\mathbf{A}(\mathbf{r}) = \int_0^\infty d\omega \sqrt{\frac{\hbar}{2\omega\epsilon_0}} \left[\mathbf{u}_\omega(\mathbf{r})\hat{a}_\omega + \mathbf{u}_\omega^*(\mathbf{r})\hat{a}_\omega^\dagger\right], \tag{S1.8}$$

with the mode functions factorized into a transverse profile and a longitudinal phase:

$$\mathbf{u}_\omega(\mathbf{r}) = \frac{\tilde{u}_\omega^{(z)}(\mathbf{r}_\perp)}{\sqrt{2\pi v_{\text{p,g}}(\omega)}} e^{iq(\omega)z}\hat{\mathbf{z}}, \tag{S1.9}$$

where $\mathbf{r}_\perp$ is the transverse coordinate, $q(\omega)$ is the dispersion relation of the photonic mode, $v_{\text{p,g}}(\omega) = (dq/d\omega)^{-1}$ is the group velocity, and $\tilde{u}_\omega^{(z)}$ is the normalized transverse profile:

$$\int d^2r_\perp \left|\tilde{u}_\omega^{(z)}(\mathbf{r}_\perp)\right|^2 = 1. \tag{S1.10}$$

Using $\int dz e^{i[q(\omega')-q(\omega)]z} = 2\pi v_{\text{p,g}}(\omega)\delta(\omega - \omega')$, the factor $\left(2\pi v_{\text{p,g}}\right)^{-1/2}$ in (S1.9) ensures the continuum orthonormality:

$$\int d^3r\, \mathbf{u}_\omega^*(\mathbf{r}) \cdot \mathbf{u}_{\omega'}(\mathbf{r}) = \delta(\omega - \omega'). \tag{S1.11}$$

## S1.3 Explicit Interaction Hamiltonian

Substituting (S1.8) and (S1.9) into (S1.5) and retaining the longitudinal component under the paraxial approximation gives:

$$H_{\text{int}} = \frac{e}{m}\int_0^\infty d\omega \sqrt{\frac{\hbar}{4\pi\omega\epsilon_0 v_{\text{p,g}}(\omega)}}\, \tilde{u}_\omega^{(\text{z})}(\mathbf{r}_\perp) e^{iq(\omega)\hat{z}} \hat{p}_z \otimes \hat{a}_\omega + \text{h.c.}, \tag{S1.12}$$

where $\hat{p}_z$ and the spatial dependence $\hat{z}$ act on the electron Hilbert space, while $\hat{a}_\omega$ and $\hat{a}_\omega^\dagger$ act on the photon Hilbert space.

## S1.4 First-Order S-Matrix Element

To first order, the transition amplitude from $|E_{\text{i}}, 0_\omega\rangle$ to $|E_{\text{f}}, 1_\omega\rangle$ is:

$$M_{E_{\text{i}}\to E_{\text{f}},\omega} = -\frac{i}{\hbar}\int_{-\infty}^{\infty} dt\, e^{-i\frac{E_{\text{i}}-E_{\text{f}}-\hbar\omega}{\hbar}t} \langle E_{\text{f}}, 1_\omega | H_{\text{int}} | E_{\text{i}}, 0_\omega\rangle. \tag{S1.13}$$

The matrix element is time independent, so the time integral yields energy conservation:

$$\int_{-\infty}^{\infty} dt\, e^{-i\frac{E_{\text{i}}-E_{\text{f}}-\hbar\omega}{\hbar}t} = 2\pi\hbar\delta(E_{\text{i}} - E_{\text{f}} - \hbar\omega). \tag{S1.14}$$

Evaluating the matrix element with the paraxial eigenstates of Eq. (S1.6) and the interaction Hamiltonian of Eq. (S1.12) gives:

$$\langle E_{\text{f}}, 1_\omega | H_{\text{int}} | E_{\text{i}}, 0_\omega\rangle = \frac{e}{m}\sqrt{\frac{\hbar}{4\pi\omega\epsilon_0 v_{\text{p,g}}}} \frac{\hbar k(E_{\text{i}})}{2\pi\hbar\sqrt{v(E_{\text{i}})v(E_{\text{f}})}}$$
$$\times \int d^2\mathbf{r}_\perp \int dz\, |\psi_\perp(\mathbf{r}_\perp)|^2 \tilde{u}_\omega^{(\text{z})*}(\mathbf{r}_\perp) e^{i[k(E_{\text{i}})-k(E_{\text{f}})-q(\omega)]z}. \tag{S1.15}$$

The integral separates when the transverse profiles $\psi_\perp$ and $\tilde{u}_\omega^{(\text{z})}$ change negligibly over $L_{\text{int}}$, i.e. when the photonic structure is translationally invariant along $z$ and $L_{\text{int}}$ is short compared with the Rayleigh range of the electron beam, so that the $z$-dependence of the integrand is carried entirely by the phase $e^{i\Delta kz}$. It then factorizes into the effective spatial overlap coefficient:

$$g_{\text{eff},\omega} = \int d^2r_\perp |\psi_\perp(\mathbf{r}_\perp)|^2 \tilde{u}_\omega^{(\text{z})*}(\mathbf{r}_\perp), \tag{S1.16}$$

which has dimensions of inverse length, and the longitudinal integral over the interaction length $L_{\text{int}}$:

$$\int_{-L_{\text{int}}/2}^{L_{\text{int}}/2} dz\, e^{i\Delta kz} = L_{\text{int}} \operatorname{sinc}\left(\Delta k \frac{L_{\text{int}}}{2}\right), \tag{S1.17}$$

where $\operatorname{sinc}(x) = \frac{\sin x}{x}$ and the phase mismatch, defined in Sec. 2 of the main text, is:

$$\Delta k(E_{\text{f}}, \omega) = k(E_{\text{f}} + \hbar\omega) - k(E_{\text{f}}) - q(\omega). \tag{S1.18}$$

Collecting the prefactors, the first-order transition matrix element is [Eq. (2) of the main text, with $v(E) \propto k(E)$ from Eq. (S1.7)]:

$$M_{E_\mathrm{i}\rightarrow E_\mathrm{f},\omega} = -i\sqrt{\frac{v(E_\mathrm{i})}{v(E_\mathrm{f})}}\sqrt{\frac{e^2}{4\pi\epsilon_0\hbar\omega v_\mathrm{p,g}}}g_{\mathrm{eff},\omega}\delta(E_\mathrm{i}-E_\mathrm{f}-\hbar\omega)L_\mathrm{int}\mathrm{sinc}\left(\frac{\Delta k(E_\mathrm{f},\omega)L_\mathrm{int}}{2}\right). \quad \text{(S1.19)}$$

## S1.5 Density Matrix of the Cooled Electron

The electron beam is described by a mixed state with density matrix $\rho_\mathrm{i}(E_\mathrm{i},E_{\mathrm{i}'})$, and the photon field is initially in the vacuum. The joint initial density operator is:

$$\rho_\mathrm{e-p}^{(\mathrm{i})} = \iint dE_\mathrm{i}dE_{\mathrm{i}'}\,\rho_\mathrm{i}(E_\mathrm{i},E_{\mathrm{i}'})|E_\mathrm{i}\rangle\langle E_{\mathrm{i}'}|\otimes|0\rangle\langle 0|, \quad \text{(S1.20)}$$

and the first-order scattering operator built from (S1.19) is:

$$\hat{S} = \int d\omega\iint dE_\mathrm{i}dE_\mathrm{f}\,M_{E_\mathrm{i}\rightarrow E_\mathrm{f},\omega}|E_\mathrm{f},1_\omega\rangle\langle E_\mathrm{i},0_\omega|, \quad \text{(S1.21)}$$

so that the final joint density operator is:

$$\rho_\mathrm{e-p}^{(\mathrm{f})} = \hat{S}\rho_\mathrm{e-p}^{(\mathrm{i})}\hat{S}^\dagger. \quad \text{(S1.22)}$$

Tracing over the photon states gives the final electron state $\rho_\mathrm{f} = \mathrm{Tr}_\mathrm{ph}\rho_\mathrm{e-p}^{(\mathrm{f})}$, and using the energy-conserving delta function to perform the $E_\mathrm{i}$ integral gives the final electron energy distribution $\rho_\mathrm{f}(E_\mathrm{f}) \equiv \rho_\mathrm{f}(E_\mathrm{f},E_\mathrm{f})$ [Eq. (3) of the main text, with $v(E)\propto k(E)$ from Eq. (S1.7)]:

$$\rho_\mathrm{f}(E_\mathrm{f}) = \frac{e^2L_\mathrm{int}^2}{4\pi\epsilon_0\hbar v_\mathrm{p,g}}\int d\omega\,\frac{v(E_\mathrm{f}+\hbar\omega)}{\omega v(E_\mathrm{f})}\left|g_{\mathrm{eff},\omega}\right|^2\mathrm{sinc}^2\left(\frac{\Delta k(E_\mathrm{f},\omega)L_\mathrm{int}}{2}\right)\rho_\mathrm{i}(E_\mathrm{f}+\hbar\omega,E_\mathrm{f}+\hbar\omega). \quad \text{(S1.23)}$$

## S1.6 Effective Coupling Strength

The total spontaneous emission probability per electron defines, to first order, the effective quantum coupling constant $g_\mathrm{Q}$:

$$P_\mathrm{em} = \int dE_\mathrm{f}\,\rho_\mathrm{f}(E_\mathrm{f}) \equiv \left|g_\mathrm{Q}\right|^2. \quad \text{(S1.24)}$$

Near the phase-matching point, the slowly varying prefactor $\frac{v(E_\mathrm{f}+\hbar\omega)}{\omega v(E_\mathrm{f})}\left|g_{\mathrm{eff},\omega}\right|^2$ of Eq. (S1.23) is evaluated at $(E_0,\omega_\mathrm{p})$. For a Gaussian initial distribution centered at $E_0$:

$$\rho_\mathrm{i}(E) = \frac{2}{\Delta E_\mathrm{i}}\sqrt{\frac{\ln 2}{\pi}}\exp\left[-4\ln 2\frac{(E-E_0)^2}{\Delta E_\mathrm{i}^2}\right], \quad \text{(S1.25)}$$

where $\rho_\mathrm{i}(E)$ denotes the diagonal element $\rho_\mathrm{i}(E,E)$, the emission probability becomes:

$$\left|g_\mathrm{Q}\right|^2 = \frac{e^2L_\mathrm{int}^2}{4\pi\epsilon_0\hbar v_\mathrm{p,g}}\frac{v(E_0)}{\omega_\mathrm{p}v(E_0-\hbar\omega_\mathrm{p})}\left|g_{\mathrm{eff},\omega_\mathrm{p}}\right|^2\int dE_\mathrm{f}\int d\omega\rho_\mathrm{i}(E_\mathrm{f}+\hbar\omega)\,\mathrm{sinc}^2\left(\Delta k(E_\mathrm{f},\omega)\frac{L_\mathrm{int}}{2}\right). \quad \text{(S1.26)}$$

## S2 Dispersion Engineering and the Cooling Conditions

### S2.1 Expansion of the Phase Mismatch

We derive Eq. (4) of the main text. With the deviation variables $\delta E_{\mathrm{f}} = E_{\mathrm{f}} - E_0 + \hbar\omega_{\mathrm{p}}$ and $\delta\omega = \omega - \omega_{\mathrm{p}}$, energy conservation gives $E_{\mathrm{i}} = E_0 + \delta E_{\mathrm{f}} + \hbar\delta\omega$, and the phase mismatch $\Delta k = k(E_{\mathrm{i}}) - k(E_{\mathrm{f}}) - q(\omega)$ becomes:

$$\Delta k(\delta E_{\mathrm{f}}, \delta\omega) = k(E_0 + \delta E_{\mathrm{f}} + \hbar\delta\omega) - k\big(E_0 - \hbar\omega_{\mathrm{p}} + \delta E_{\mathrm{f}}\big) - q\big(\omega_{\mathrm{p}} + \delta\omega\big), \tag{S2.1}$$

which vanishes at the phase-matching point $\delta E_{\mathrm{f}} = \delta\omega = 0$. Using $\partial k/\partial E = 1/(\hbar v)$ and $\partial q/\partial\omega = 1/v_{\mathrm{p,g}}$, the first derivatives at this point are:

$$\frac{\partial\Delta k}{\partial\delta E_{\mathrm{f}}} = \left.\frac{\partial k}{\partial E}\right|_{E_0} - \left.\frac{\partial k}{\partial E}\right|_{E_0-\hbar\omega_{\mathrm{p}}} = \frac{\Delta v_{\mathrm{i-f}}^{-1}}{\hbar}, \quad \frac{\partial\Delta k}{\partial\delta\omega} = \hbar\left.\frac{\partial k}{\partial E}\right|_{E_0} - \left.\frac{\partial q}{\partial\omega}\right|_{\omega_{\mathrm{p}}} = \Delta v_{\mathrm{GVM}}^{-1}, \tag{S2.2}$$

and the second derivatives are:

$$\frac{\partial^2\Delta k}{\partial\delta E_{\mathrm{f}}^2} = \left.\frac{\partial^2 k}{\partial E^2}\right|_{E_0} - \left.\frac{\partial^2 k}{\partial E^2}\right|_{E_0-\hbar\omega_{\mathrm{p}}}, \quad \frac{\partial^2\Delta k}{\partial\delta E_{\mathrm{f}}\partial\delta\omega} = \hbar\left.\frac{\partial^2 k}{\partial E^2}\right|_{E_0},$$
$$\frac{\partial^2\Delta k}{\partial\delta\omega^2} = \hbar^2\left.\frac{\partial^2 k}{\partial E^2}\right|_{E_0} - \left.\frac{\partial^2 q}{\partial\omega^2}\right|_{\omega_{\mathrm{p}}}. \tag{S2.3}$$

Collecting terms reproduces Eq. (4) of the main text, here including the term of second order in $\delta E_{\mathrm{f}}$:

$$\Delta k(\delta E_{\mathrm{f}}, \delta\omega) \approx \left[\frac{\Delta v_{\mathrm{i-f}}^{-1}}{\hbar} + \hbar\left.\frac{\partial^2 k}{\partial E^2}\right|_{E_0}\delta\omega\right]\delta E_{\mathrm{f}} + \Delta v_{\mathrm{GVM}}^{-1}\delta\omega$$
$$+\frac{1}{2}\left[\hbar^2\left.\frac{\partial^2 k}{\partial E^2}\right|_{E_0} - \left.\frac{\partial^2 q}{\partial\omega^2}\right|_{\omega_{\mathrm{p}}}\right]\delta\omega^2 + \frac{1}{2}\left[\left.\frac{\partial^2 k}{\partial E^2}\right|_{E_0} - \left.\frac{\partial^2 k}{\partial E^2}\right|_{E_0-\hbar\omega_{\mathrm{p}}}\right]\delta E_{\mathrm{f}}^2. \tag{S2.4}$$

### S2.2 Dispersion-Engineering Conditions and Validity of the Expansion

In addition to phase matching, optimal cooling requires $v_{\mathrm{p,g}} = v_{\mathrm{e}}$ and $\left.\frac{\partial^2 q}{\partial\omega^2}\right|_{\omega_{\mathrm{p}}} = \hbar^2\left.\frac{\partial^2 k}{\partial E^2}\right|_{E_0}$ (Sec. 2 of the main text). For a non-relativistic electron: $k(E) = \sqrt{2Em}/\hbar$ and $v = \sqrt{2E/m}$, such that the initial and final electron velocities are $v_{\mathrm{e}} = \sqrt{2E_0/m}$ and $v_{\mathrm{e,f}} = \sqrt{2E_0/m - \frac{4\pi\hbar c}{m\lambda_{\mathrm{p}}}} = \sqrt{v_{\mathrm{e}}^2 - \frac{4\pi\hbar c}{m\lambda_{\mathrm{p}}}}$, therefore $\Delta v_{\mathrm{i-f}}^{-1} = v_{\mathrm{e}}^{-1} - \left(v_{\mathrm{e}}^2 - \frac{4\pi\hbar c}{m\lambda_{\mathrm{p}}}\right)^{-1/2}$. Inserting this into Eq. (8a) of the main text gives the optimal cooling ratio $\frac{\Delta E_{\mathrm{f}}}{\Delta E_{\mathrm{i}}} \simeq \frac{2\mu\hbar}{L_{\mathrm{int}}\left[\left(v_{\mathrm{e}}^2 - \frac{4\pi\hbar c}{m\lambda_{\mathrm{p}}}\right)^{-1/2} - v_{\mathrm{e}}^{-1}\right]\Delta E_{\mathrm{i}}}$, where $\mu = 2.783$ is the FWHM of $\mathrm{sinc}^2 x$, and the threshold interaction length after which cooling begins is $L_0 = \frac{2\mu\hbar}{\left[\left(v_{\mathrm{e}}^2 - \frac{4\pi\hbar c}{m\lambda_{\mathrm{p}}}\right)^{-1/2} - v_{\mathrm{e}}^{-1}\right]\Delta E_{\mathrm{i}}}$. In the weak-recoil limit $\hbar\omega_{\mathrm{p}} \ll E_0$, this reduces to $\frac{L_0}{\lambda_{\mathrm{p}}} \simeq \frac{\mu m v_{\mathrm{e}}^3}{\pi c\Delta E_{\mathrm{i}}}$. The electron's dispersion satisfies $\partial^2 k/\partial E^2 = -m^2/(\hbar^4 k^3) < 0$ at all energies, in both the non-relativistic and the relativistic regime, so the required group-velocity dispersion of the photonic mode is the same, $\left.\frac{\partial^2 q}{\partial\omega^2}\right|_{\omega_{\mathrm{p}}} < 0$, for any electron energy.

Terms of order $\delta E_{\mathrm{f}}^2$ in Eq. (S2.4) are neglected relative to the linear term. For the non-relativistic dispersion $k(E)=\sqrt{2Em}/\hbar$, $\partial^2 k/\partial E^2=-(\partial k/\partial E)/2E$, and the ratio of the two terms is:

$$\frac{\left|\frac{1}{2}\left[\frac{\partial^2 k}{\partial E^2}\Big|_{\mathrm{E_0}}-\frac{\partial^2 k}{\partial E^2}\Big|_{\mathrm{E_0}-\hbar\omega_{\mathrm{p}}}\right]\delta E_{\mathrm{f}}^2\right|}{\left|\Delta v_{\mathrm{i-f}}^{-1}\delta E_{\mathrm{f}}/\hbar\right|}=\frac{|\delta E_{\mathrm{f}}|}{4}\left(\frac{1}{E_0}+\frac{1}{\sqrt{E_0(E_0-\hbar\omega_{\mathrm{p}})}}+\frac{1}{E_0-\hbar\omega_{\mathrm{p}}}\right)\leq\frac{3|\delta E_{\mathrm{f}}|}{4(E_0-\hbar\omega_{\mathrm{p}})},\qquad(\mathrm{S2.5})$$

which is small since $|\delta E_{\mathrm{f}}|\lesssim\Delta E_{\mathrm{i}}\ll E_0-\hbar\omega_{\mathrm{p}}$. For $\Delta E_{\mathrm{i}}=0.584$ eV and $v_{\mathrm{e}}=0.1c$, it is about $2\cdot10^{-4}$.

Eq. (S2.4) is truncated at second order in $\delta\omega$. At $v_{\mathrm{p,g}}=v_{\mathrm{e}}$, the leading neglected term, $\frac{1}{6}\left[\hbar^3\frac{\partial^3 k}{\partial E^3}\Big|_{E_0}-\frac{\partial^3 q}{\partial\omega^3}\Big|_{\omega_{\mathrm{p}}}\right]\delta\omega^3$, must not shift the phase-matching point by more than the phase-matching window over the emitted band $\hbar|\delta\omega|\lesssim\Delta E_{\mathrm{i}}$, so the truncation holds when:

$$\left|\hbar^3\frac{\partial^3 k}{\partial E^3}\Big|_{E_0}-\frac{\partial^3 q}{\partial\omega^3}\Big|_{\omega_{\mathrm{p}}}\right|\left(\frac{\Delta E_{\mathrm{i}}}{\hbar}\right)^3 L_{\mathrm{int}}\ll 1,\qquad(\mathrm{S2.6})$$

i.e. the residual cubic phase accumulated over the interaction length must stay within the phase-matching window. Unlike the window, this residual cubic phase grows with $L_{\mathrm{int}}$, so at long interaction lengths it limits $\Delta E_{\mathrm{f}}$ unless the third-order dispersion is also engineered.

## S2.3 Group-Velocity Mismatch: Cooling and Heating

This section supplements Eqs. (5)–(8) of the main text. The heating resonance at $v_{\mathrm{p,g}}=v_{\mathrm{e,f}}$ can be seen by using energy conservation, $\delta E_{\mathrm{i}}=\delta E_{\mathrm{f}}+\hbar\delta\omega$, to eliminate $\delta\omega$ from the linear terms of the phase mismatch, Eq. (4) of the main text, which gives $\Delta k\approx\left[\left(v_{\mathrm{p,g}}^{-1}-v_{\mathrm{e,f}}^{-1}\right)\delta E_{\mathrm{f}}+\Delta v_{\mathrm{GVM}}^{-1}\delta E_{\mathrm{i}}\right]/\hbar$. Phase matching, $\Delta k=0$, then fixes $\delta E_{\mathrm{f}}$ in terms of $\delta E_{\mathrm{i}}$, which is Eq. (6) of the main text. At $v_{\mathrm{p,g}}=v_{\mathrm{e,f}}$ the coefficient of $\delta E_{\mathrm{f}}$ vanishes, so $\Delta k\approx\Delta v_{\mathrm{i-f}}^{-1}\delta E_{\mathrm{i}}/\hbar$ no longer depends on the final energy: phase matching selects the initial energy but leaves $\delta E_{\mathrm{f}}$ unconstrained, and the linear theory predicts an unbounded final spread, the divergence of Eq. (6). In practice, the spread stays finite because of the terms neglected in Eq. (6) of the main text, namely the mixed term $\hbar\frac{\partial^2 k}{\partial E^2}\Big|_{E_0}\delta\omega\delta E_{\mathrm{f}}$ and the $\delta E_{\mathrm{f}}^2$ term of Eq. (S2.4), and because of the finite photonic bandwidth, which limits the emitted photon frequencies (Sec. S4.5). Together they keep $\Delta E_{\mathrm{f}}$ finite. The heatmap of Fig. 3b of the main text, computed from the full Eq. (3), includes these contributions.

Away from this resonance, the phase-matching window maps onto a final-energy window of width $2\mu\hbar/\left(L_{\mathrm{int}}\left|v_{\mathrm{p,g}}^{-1}-v_{\mathrm{e,f}}^{-1}\right|\right)$, centered on the value of $\delta E_{\mathrm{f}}$ given by Eq. (6) of the main text. The final FWHM therefore combines the mapped initial spread with this window, which we estimate by:

$$\Delta E_{\mathrm{f}}\approx\frac{\left|v_{\mathrm{e}}^{-1}-v_{\mathrm{p,g}}^{-1}\right|\Delta E_{\mathrm{i}}+2\mu\hbar/L_{\mathrm{int}}}{\left|v_{\mathrm{e,f}}^{-1}-v_{\mathrm{p,g}}^{-1}\right|},\qquad(\mathrm{S2.7})$$

which reduces to Eq. (8a) of the main text at $v_{\mathrm{p,g}}=v_{\mathrm{e}}$ and to the $L_{\mathrm{int}}$-independent value of Eq. (6) of the main text at long interaction lengths (the linear sum is an upper estimate of the convolution width).

At long interaction lengths, the window term of Eq. (S2.7) is negligible. Near group-velocity matching, $|v_{\mathrm{p,g}}-v_{\mathrm{e}}|\ll v_{\mathrm{e}}-v_{\mathrm{e,f}}$, its numerator is $\approx|v_{\mathrm{p,g}}-v_{\mathrm{e}}|\Delta E_{\mathrm{i}}/v_{\mathrm{e}}^2$ and its denominator is $\approx(v_{\mathrm{e}}-v_{\mathrm{e,f}})/v_{\mathrm{e}}^2$, so that

$\Delta E_\mathrm{f}/\Delta E_\mathrm{i} \approx |v_\mathrm{p,g} - v_\mathrm{e}|/(v_\mathrm{e} - v_\mathrm{e,f})$. A target cooling ratio therefore requires the group velocity tolerance of (the last step is non-relativistic):

$$\frac{|v_\mathrm{p,g} - v_\mathrm{e}|}{v_\mathrm{e}} \lesssim \frac{\Delta E_\mathrm{f}}{\Delta E_\mathrm{i}} \frac{v_\mathrm{e} - v_\mathrm{e,f}}{v_\mathrm{e}} \approx \frac{\Delta E_\mathrm{f}}{\Delta E_\mathrm{i}} \frac{\hbar\omega_\mathrm{p}}{2E_0}. \tag{S2.8}$$

# S3 Green's Function Approach

The guided-mode treatment of Sec. S1 assumes an ideal, lossless photonic mode. Here we treat the lossy photonic structure within macroscopic quantum electrodynamics (MQED) [8,9], as in Sec. 2 of the main text, derive the lossy final electron energy distribution of Eq. (9) of the main text, and show that it reduces to Eq. (3) of the main text in the lossless limit.

## S3.1 MQED Field Quantization

In MQED, the vector potential is expressed through the dyadic Green's function $\mathbf{G}(\mathbf{r}, \mathbf{r}', \omega)$ of the photonic structure, which gives the field at $\mathbf{r}$ generated by a unit dipole at $\mathbf{r}'$, and bosonic noise-field operators $\hat{\mathbf{f}}(\mathbf{r}, \omega)$:

$$\hat{\mathbf{A}}(\mathbf{r}, t) = \sqrt{\frac{\hbar}{\pi\epsilon_0}} \int_0^\infty d\omega \, \frac{\omega}{c^2} \int d^3\mathbf{r}' \sqrt{\mathrm{Im}\,\varepsilon(\mathbf{r}', \omega)} \mathbf{G}(\mathbf{r}, \mathbf{r}', \omega) \hat{\mathbf{f}}(\mathbf{r}', \omega) e^{-i\omega t} + \mathrm{h.c.}, \tag{S3.1}$$

where $\varepsilon(\mathbf{r}, \omega)$ is the complex permittivity and the noise operators satisfy:

$$[\hat{f}_i(\mathbf{r}, \omega), \hat{f}_j^\dagger(\mathbf{r}', \omega')] = \delta_{ij}\delta(\omega - \omega')\delta^3(\mathbf{r} - \mathbf{r}'). \tag{S3.2}$$

The weight $\sqrt{\mathrm{Im}\,\varepsilon(\mathbf{r}', \omega)}$ expresses the fluctuation–dissipation theorem: every volume element with finite $\mathrm{Im}\,\varepsilon$ acts as a quantum noise source. The Green's function satisfies the Helmholtz equation:

$$\nabla \times \nabla \times \mathbf{G}(\mathbf{r}, \mathbf{r}', \omega) - \frac{\omega^2}{c^2} \varepsilon(\mathbf{r}, \omega) \mathbf{G}(\mathbf{r}, \mathbf{r}', \omega) = \delta^3(\mathbf{r} - \mathbf{r}')\mathbf{I}, \tag{S3.3}$$

and the optical theorem:

$$\frac{\omega^2}{c^2} \sum_k \int d^3\mathbf{s} \, \mathrm{Im}\,\varepsilon(\mathbf{s}, \omega) G_{ik}(\mathbf{r}, \mathbf{s}, \omega) G_{jk}^*(\mathbf{r}', \mathbf{s}, \omega) = \mathrm{Im}\, G_{ij}(\mathbf{r}, \mathbf{r}', \omega), \tag{S3.4}$$

which states that the imaginary part of the Green's function equals the power absorbed by all dissipation channels. It is used in Sec. S3.4 to eliminate the sum over noise sources.

## S3.2 Transition Amplitude

The initial and final electron–field states are $|\Psi_\mathrm{i}\rangle = |E_\mathrm{i}\rangle \otimes |0\rangle$ and $|\Psi_\mathrm{f}\rangle = |E_\mathrm{f}\rangle \otimes |1_{\mathbf{r}_0, j, \omega}\rangle$, where $|1_{\mathbf{r}_0, j, \omega}\rangle$ is a single MQED excitation at position $\mathbf{r}_0$, with polarization $j$ and frequency $\omega$. With the interaction Hamiltonian $(e/m)\hat{A}_z\hat{p}_z$ of Sec. S1.3, the first-order transition amplitude is:

$$M_{E_\mathrm{i}\to E_\mathrm{f}, \omega}(\mathbf{r}_0, j) = -\frac{i}{\hbar}\frac{e}{m} \int_{-\infty}^{\infty} dt \, \langle \Psi_\mathrm{f} | \hat{A}_z(\mathbf{r}, t)\hat{p}_z | \Psi_\mathrm{i} \rangle. \tag{S3.5}$$

For spontaneous emission into the field vacuum, only the creation part of Eq. (S3.1) contributes, and Eq. (S3.1) gives:

$$\langle 1_{\mathbf{r}_0,j,\omega}|\hat{A}_z(\mathbf{r},t)|0\rangle = \sqrt{\frac{\hbar}{\pi\epsilon_0}\frac{\omega}{c^2}}\sqrt{\mathrm{Im}\,\varepsilon(\mathbf{r}_0,\omega)}G^*_{zj}(\mathbf{r},\mathbf{r}_0,\omega)e^{i\omega t}. \tag{S3.6}$$

The electron states are the energy-normalized paraxial states of Eq. (S1.6), $\langle\mathbf{r}|E\rangle = \psi_\perp(\mathbf{r}_\perp)e^{ik(E)z}/\sqrt{2\pi\hbar v(E)}$, with $\hat{p}_z|E\rangle = \hbar k(E)|E\rangle$ and $\hbar k(E)/m = v(E)$ for low-energy electrons. In the paraxial narrow-beam regime, the Green's function varies slowly across the beam and is evaluated at the beam center $\mathbf{r}_{\mathrm{T,e}}$, i.e. along the electron trajectory $\mathbf{r}_z \equiv \mathbf{r}_{\mathrm{T,e}} + z\hat{\mathbf{z}}$. The time integral, $\int dt\, e^{i(E_\mathrm{f}+\hbar\omega-E_\mathrm{i})t/\hbar} = 2\pi\hbar\delta(E_\mathrm{i}-E_\mathrm{f}-\hbar\omega)$, enforces energy conservation, and the amplitude becomes:

$$M_{E_\mathrm{i}\to E_\mathrm{f},\omega}(\mathbf{r}_0,j) = -i\frac{e\omega}{c^2}\sqrt{\frac{v_\mathrm{e}}{\pi\epsilon_0\hbar v_{\mathrm{e,f}}}}\sqrt{\mathrm{Im}\,\varepsilon(\mathbf{r}_0,\omega)}$$
$$\times\int_{-\infty}^{\infty} dz\, e^{i[k(E_\mathrm{i})-k(E_\mathrm{f})]z}G^*_{zj}(\mathbf{r}_z,\mathbf{r}_0,\omega)\,\delta(E_\mathrm{i}-E_\mathrm{f}-\hbar\omega). \tag{S3.7}$$

## S3.3 Joint Density Matrix and Photon Trace

For an initial electron density matrix $\rho_\mathrm{i}(E_\mathrm{i},E'_\mathrm{i})$ and the field in its vacuum state, the electron–field density operator after the interaction is composed of $M_{E_\mathrm{i}\to E_\mathrm{f},\omega}(\mathbf{r}_0,j)$ and its complex conjugate. Integrating over $E_\mathrm{i}$ and $E'_\mathrm{i}$ with the energy-conserving delta functions, which set $E_\mathrm{i} = E_\mathrm{f} + \hbar\omega$, and tracing over the photon states, which sets $\omega' = \omega$, $\mathbf{r}'_0 = \mathbf{r}_0$ and $j' = j$, gives the reduced electron density matrix:

$$\rho_\mathrm{f} = \frac{e^2}{\pi\hbar\epsilon_0 c^2}\sum_j\int d^3\mathbf{r}_0\int d\omega\int dE_\mathrm{f}\int dE'_\mathrm{f}\sqrt{\frac{v(E_\mathrm{f}+\hbar\omega)v(E'_\mathrm{f}+\hbar\omega)}{v(E_\mathrm{f})v(E'_\mathrm{f})}}\frac{\omega^2}{c^2}\mathrm{Im}\,\varepsilon(\mathbf{r}_0,\omega)$$
$$\times\int dz\, e^{i[k(E_\mathrm{f}+\hbar\omega)-k(E_\mathrm{f})]z}G^*_{zj}(\mathbf{r}_z,\mathbf{r}_0,\omega)\int dz'\, e^{-i[k(E'_\mathrm{f}+\hbar\omega)-k(E'_\mathrm{f})]z'}G_{zj}(\mathbf{r}_{z'},\mathbf{r}_0,\omega)$$
$$\times\,\rho_\mathrm{i}(E_\mathrm{f}+\hbar\omega,E'_\mathrm{f}+\hbar\omega)|E_\mathrm{f}\rangle\langle E'_\mathrm{f}|. \tag{S3.8}$$

## S3.4 Applying the Optical Theorem

The sum over $j$ and the integral over $\mathbf{r}_0$ in Eq. (S3.8) have the form of the optical theorem, Eq. (S3.4), with $\mathbf{r} = \mathbf{r}_z$ and $\mathbf{r}' = \mathbf{r}_{z'}$:

$$\frac{\omega^2}{c^2}\sum_j\int d^3\mathbf{r}_0\,\mathrm{Im}\,\varepsilon(\mathbf{r}_0,\omega)G^*_{zj}(\mathbf{r}_z,\mathbf{r}_0,\omega)G_{zj}(\mathbf{r}_{z'},\mathbf{r}_0,\omega) = \mathrm{Im}\,G_{zz}(\mathbf{r}_z,\mathbf{r}_{z'},\omega), \tag{S3.9}$$

such that the diagonal element of Eq. (S3.8), $E'_\mathrm{f} = E_\mathrm{f}$, gives the general MQED result for the final electron energy distribution:

$$\rho_\mathrm{f}(E_\mathrm{f}) = \frac{e^2}{\pi\hbar\epsilon_0 c^2}\int d\omega\,\frac{v(E_\mathrm{f}+\hbar\omega)}{v(E_\mathrm{f})}\rho_\mathrm{i}(E_\mathrm{f}+\hbar\omega,E_\mathrm{f}+\hbar\omega)$$
$$\times\int dz\int dz'\, e^{i[k(E_\mathrm{f}+\hbar\omega)-k(E_\mathrm{f})](z-z')}\mathrm{Im}\,G_{zz}(\mathbf{r}_z,\mathbf{r}_{z'},\omega)\,. \tag{S3.10}$$

## S3.5 Guided Mode, Cavity Mode and the Continuum Limit

To reach the guided-mode result for Eq. (S3.10), consider first a cavity of length $L_\mathrm{int}$ and cross-section $A$ (volume $V = AL_\mathrm{int}$) with modes $\mathbf{u}_n(\mathbf{r})$ of frequency $\omega_n$ and decay rate $\alpha_n$. Only the transverse part of its Green's function is resonant with the photonic modes. The longitudinal part is quasi-static, and for a non-absorbing structure it is real, so that it does not contribute to the imaginary part of the Green's function in Eq. (S3.10). For weak absorption, it adds a non-radiative near-field channel, proportional to the imaginary part of the permittivity, which is not emission into the mode and is neglected here [10]. The Green's function in Eq. (S3.10) [11] is therefore replaced by its transverse part, expanded in the cavity modes [12]:

$$\mathbf{G}^{\mathrm{T}}(\mathbf{r},\mathbf{r}',\omega)=\sum_n \mathbf{u}_n(\mathbf{r})\mathbf{u}_n^*(\mathbf{r}')\frac{c^2}{2\omega_n}\left[\frac{1}{\omega_n-\omega-i\alpha_n/2}+\frac{1}{\omega_n+\omega+i\alpha_n/2}\right], \tag{S3.11}$$

and keeping the resonant term:

$$\operatorname{Im} G_{zz}^{\mathrm{T}}(\mathbf{r}_z,\mathbf{r}_{z'},\omega)=\sum_n \frac{\left|u_{n,z}(\mathbf{r}_{\mathrm{T,e}})\right|^2}{V}u_n(z)u_n^*(z')\frac{c^2}{2\omega_n}\frac{\alpha_n/2}{(\omega_n-\omega)^2+(\alpha_n/2)^2}. \tag{S3.12}$$

A cavity of length $L_{\mathrm{int}}$ with reflecting ends supports real standing-wave modes, $u_n(z)=\sqrt{2}\cos(q_n z)$, normalized so that $\int dz\,|u_n(z)|^2=L_{\mathrm{int}}$. Each mode is the sum of two counter-propagating travelling waves, $\sqrt{2}\cos(q_n z)=\left(e^{iq_n z}+e^{-iq_n z}\right)/\sqrt{2}$. In the longitudinal integral we therefore set $u_n(z)\to e^{-iq_n z}/\sqrt{2}$ for $|z|<L_{\mathrm{int}}/2$, keeping only the travelling wave that is phase-matched to the electron (the other component, $e^{iq_n z}/\sqrt{2}$, has a phase mismatch of about $2q_n\gg 1/L_{\mathrm{int}}$ and averages out). The longitudinal integral is then:

$$\left|\int_{-L_{\mathrm{int}}/2}^{L_{\mathrm{int}}/2} dz\, e^{i[k(E_{\mathrm{f}}+\hbar\omega)-k(E_{\mathrm{f}})]z}u_n(z)\right|^2=\frac{L_{\mathrm{int}}^2}{2}\operatorname{sinc}^2\left(\frac{\Delta k_n L_{\mathrm{int}}}{2}\right), \tag{S3.13}$$

where $\Delta k_n=k(E_{\mathrm{f}}+\hbar\omega)-k(E_{\mathrm{f}})-q_n$ is the phase mismatch of mode $n$. Substituting Eqs. (S3.12) and (S3.13) into Eq. (S3.10), taking the continuum limit $\sum_n \to \left(L_{\mathrm{int}}/\pi v_{\mathrm{p,g}}\right)\int d\omega'$ for the standing-wave mode spacing $\pi v_{\mathrm{p,g}}/L_{\mathrm{int}}$ with $\omega_n\to\omega'$, $q_n\to q(\omega')$, $\alpha_n\to\alpha$, and averaging over the transverse electron wavefunction (replacing $\left|u_{n,z}(\mathbf{r}_{\mathrm{T,e}})\right|^2/A$ by $\left|g_{\mathrm{eff},\omega'}\right|^2$), gives:

$$\begin{aligned}\rho_{\mathrm{f}}(E_{\mathrm{f}})=\frac{e^2L_{\mathrm{int}}^2}{4\pi\epsilon_0\hbar v_{\mathrm{p,g}}}\int d\omega\,\frac{v(E_{\mathrm{f}}+\hbar\omega)}{v(E_{\mathrm{f}})}\rho_{\mathrm{i}}(E_{\mathrm{f}}+\hbar\omega,E_{\mathrm{f}}+\hbar\omega)\\ \times\int d\omega'\,\frac{\left|g_{\mathrm{eff},\omega'}\right|^2}{\omega'}\operatorname{sinc}^2\left(\frac{\Delta k(E_{\mathrm{f}},\omega')L_{\mathrm{int}}}{2}\right)\frac{1}{\pi}\frac{\alpha/2}{(\omega'-\omega)^2+(\alpha/2)^2},\end{aligned} \tag{S3.14}$$

where $\Delta k(E_{\mathrm{f}},\omega')=k(E_{\mathrm{f}}+\hbar\omega)-k(E_{\mathrm{f}})-q(\omega')$. This is the lossy distribution of Eq. (9) of the main text, with $v(E)\propto k(E)$ from Eq. (S1.7). For a narrow beam, $\left|g_{\mathrm{eff},\omega'}\right|^2$ reduces to the point value $\left|\tilde{u}_{\omega'}^{(z)}(\mathbf{r}_{\mathrm{T,e}})\right|^2$. As $\alpha\to 0$ the Lorentzian tends to $\delta(\omega'-\omega)$, and Eq. (S3.14) reduces to:

$$\rho_{\mathrm{f}}(E_{\mathrm{f}})=\frac{e^2L_{\mathrm{int}}^2}{4\pi\epsilon_0\hbar v_{\mathrm{p,g}}}\int d\omega\,\frac{v(E_{\mathrm{f}}+\hbar\omega)}{\omega v(E_{\mathrm{f}})}\left|g_{\mathrm{eff},\omega}\right|^2\operatorname{sinc}^2\left(\frac{\Delta k(E_{\mathrm{f}},\omega)L_{\mathrm{int}}}{2}\right)\rho_{\mathrm{i}}(E_{\mathrm{f}}+\hbar\omega,E_{\mathrm{f}}+\hbar\omega), \tag{S3.15}$$

which is Eq. (S1.23) and has the form of Eq. (3) of the main text. For $\alpha>0$, the $\omega'$ integral is a convolution of the $\operatorname{sinc}^2$ kernel with the Lorentzian lineshape of the photonic mode. As $L_{\mathrm{int}}\to\infty$ the kernel narrows while the Lorentzian keeps its width $\alpha$, which sets the saturation floor derived in Sec. S4.

## S4 Saturation Due to Photonic Losses

Photonic loss imposes a lower limit on the attainable energy spread (Sec. 2 of the main text). This section derives the loss-limited floor of Eq. (10) of the main text from the lossy distribution of Eq. (9).

### S4.1 Modified Density Matrix with Lorentzian Broadening

Eq. (9) of the main text follows from the Green's-function treatment of Sec. S3 when the finite decay rate of the photonic mode is retained in the cavity Green's function, Eq. (S3.11). Each mode frequency is then broadened into the normalized Lorentzian:

$$\mathcal{L}(x) = \frac{1}{\pi}\frac{\alpha/2}{x^2 + (\alpha/2)^2}, \tag{S4.1}$$

where $x = \omega' - \omega$ is the detuning from the mode frequency and $\alpha$, the temporal intensity loss rate, is the FWHM of $\mathcal{L}(x)$. At the engineered point ($v_{\mathrm{p,g}} = v_{\mathrm{e}}$ and $\left.\frac{\partial^2 q}{\partial \omega^2}\right|_{\omega_{\mathrm{p}}} = \hbar^2 \left.\frac{\partial^2 k}{\partial E^2}\right|_{E_0}$, Sec. 2 of the main text), the phase mismatch at the mode frequency $\omega'$ is $\Delta k \approx \Delta v_{\mathrm{i-f}}^{-1} \delta E_{\mathrm{f}}/\hbar - x/v_{\mathrm{p,g}}$, and Eq. (9) of the main text becomes, in the deviation variables of Sec. S2.1:

$$\rho_{\mathrm{f}}(\delta E_{\mathrm{f}}) \propto \int d\delta\omega\, \rho_{\mathrm{i}}(E_0 + \delta E_{\mathrm{f}} + \hbar\delta\omega) \int dx\, \mathrm{sinc}^2\left[\left(\frac{\Delta v_{\mathrm{i-f}}^{-1}}{\hbar}\delta E_{\mathrm{f}} - \frac{x}{v_{\mathrm{p,g}}}\right)\frac{L_{\mathrm{int}}}{2}\right]\mathcal{L}(x). \tag{S4.2}$$

### S4.2 Saturation Limit

The $\delta\omega$ integral involves only the initial distribution and gives a constant, so the final distribution is the convolution of the $\mathrm{sinc}^2$ kernel with the Lorentzian:

$$\rho_{\mathrm{f}}(\delta E_{\mathrm{f}}) \propto \int dx\, \mathrm{sinc}^2\left[\left(\frac{\Delta v_{\mathrm{i-f}}^{-1}}{\hbar}\delta E_{\mathrm{f}} - \frac{x}{v_{\mathrm{p,g}}}\right)\frac{L_{\mathrm{int}}}{2}\right]\mathcal{L}(x). \tag{S4.3}$$

In $x$, the kernel has FWHM $2\mu v_{\mathrm{p,g}}/L_{\mathrm{int}}$ and the Lorentzian has FWHM $\alpha$. For $L_{\mathrm{int}} \gg v_{\mathrm{p,g}}/\alpha$ the kernel approaches $(2\pi v_{\mathrm{p,g}}/L_{\mathrm{int}})\delta(x - v_{\mathrm{p,g}}\Delta v_{\mathrm{i-f}}^{-1}\delta E_{\mathrm{f}}/\hbar)$, and the final distribution becomes a Lorentzian in $\delta E_{\mathrm{f}}$:

$$\rho_{\mathrm{f}}(\delta E_{\mathrm{f}}) \propto \mathcal{L}\left(\frac{v_{\mathrm{p,g}}\Delta v_{\mathrm{i-f}}^{-1}}{\hbar}\delta E_{\mathrm{f}}\right), \tag{S4.4}$$

whose FWHM is the saturation floor given in Sec. 2 of the main text (the last two forms are non-relativistic):

$$\Delta E_{\mathrm{f,sat}} = \frac{\hbar\alpha}{v_{\mathrm{p,g}}\left|\Delta v_{\mathrm{i-f}}^{-1}\right|} \approx \frac{2 v_{\mathrm{e}} E_0 \alpha}{\omega_{\mathrm{p}} v_{\mathrm{p,g}}} = \frac{v_{\mathrm{e}} E_0 \alpha \lambda_{\mathrm{p}}}{\pi c v_{\mathrm{p,g}}}. \tag{S4.5}$$

The floor is independent of $\Delta E_{\mathrm{i}}$ and $L_{\mathrm{int}}$ and depends only on the spatial attenuation coefficient $\alpha/v_{\mathrm{p,g}}$. At $v_{\mathrm{p,g}} = v_{\mathrm{e}}$, relative to the beam energy and for non-relativistic electrons, the floor is twice the inverse quality factor of the photonic mode [Eq. (10) of the main text]:

$$\frac{\Delta E_{\mathrm{f,sat}}}{E_0} \simeq \frac{2\alpha}{\omega_{\mathrm{p}}}. \tag{S4.6}$$

## S4.3 Convergence to the Lossless Limit

As $\alpha \to 0^+$, the Lorentzian of Eq. (S4.1) is non-negative, normalized to unity and concentrates at $x = 0$, so that:

$$\mathcal{L}(x) \to \delta(x) \quad (\alpha \to 0^+). \tag{S4.7}$$

Eq. (S4.3) then reduces to the lossless kernel $\mathrm{sinc}^2\left(\Delta v_{\mathrm{i-f}}^{-1} L_{\mathrm{int}} \delta E_{\mathrm{f}}/2\hbar\right)$ of Eq. (3) of the main text, and Eq. (S4.5) gives $\Delta E_{\mathrm{f,sat}} \to 0$: the lossless result, $\Delta E_{\mathrm{f}} \propto 1/L_{\mathrm{int}}$ at all interaction lengths, is recovered.

## S4.4 Crossover and Saturation Length

For $L_{\mathrm{int}} \ll v_{\mathrm{p,g}}/\alpha$, the Lorentzian is narrower than the kernel, and at $v_{\mathrm{p,g}} = v_{\mathrm{e}}$ the final width is the phase-matching window of Sec. 2 of the main text, $\Delta E_{\mathrm{f}}^{\mathrm{win}} = 2\mu\hbar/\left(L_{\mathrm{int}}\left|\Delta v_{\mathrm{i-f}}^{-1}\right|\right)$. Equating the two widths gives the crossover interaction length:

$$L_{\mathrm{int}} \approx \frac{2\mu v_{\mathrm{p,g}}}{\alpha}, \tag{S4.8}$$

i.e. $2\mu$ times the attenuation length $v_{\mathrm{p,g}}/\alpha$ that limits $L_{\mathrm{loss}}$ in Sec. 2 of the main text. The two regimes are interpolated:

$$\Delta E_{\mathrm{f}} \approx \frac{\hbar}{\left|\Delta v_{\mathrm{i-f}}^{-1}\right|} \max\left\{\frac{2\mu}{L_{\mathrm{int}}}, \frac{\alpha}{v_{\mathrm{p,g}}}\right\}, \tag{S4.9}$$

while the full finite-length width, including the initial distribution, is obtained numerically from Eq. (9) of the main text (Fig. S1a and Fig. 3a of the main text). Beyond this length, increasing $L_{\mathrm{int}}$ no longer narrows the electron distribution.

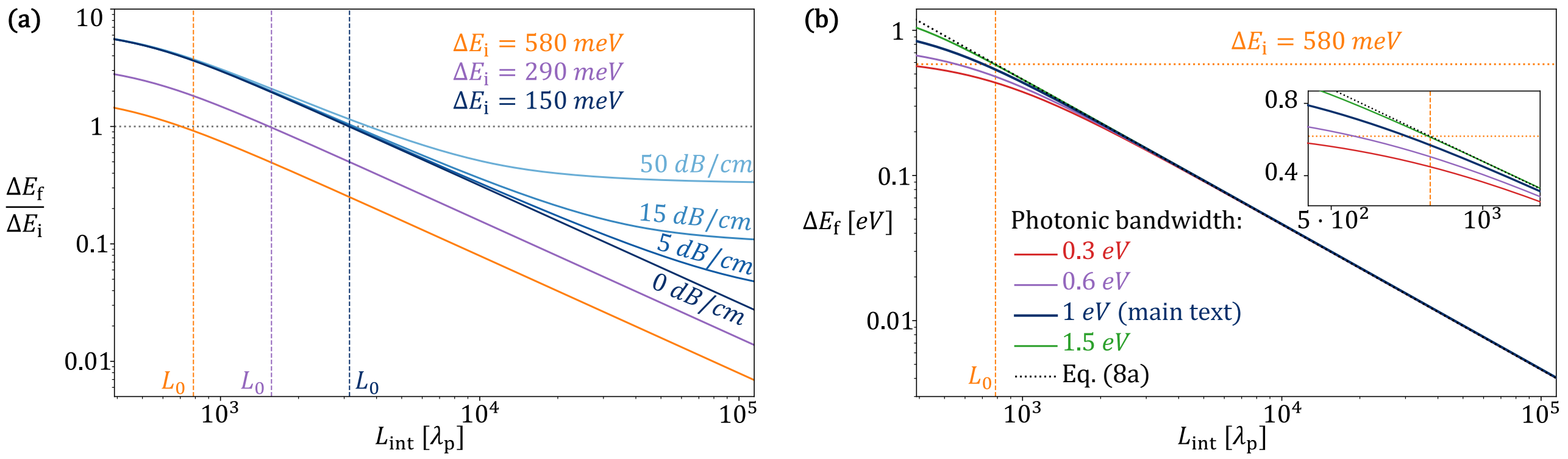

**Fig. S1: Finite interaction-length and photonic-loss limits. (a)** Cooling ratio $\Delta E_f/\Delta E_i$ as a function of $L_{int}/\lambda_p$ at $v_e = v_{p,g} = 0.1c$ and $\lambda_p = 500$ nm, where $\Delta E_f$ and $\Delta E_i$ are the final and initial electron energy spreads (FWHM). The lossless curves are shown for $\Delta E_i = 0.584$, 0.292 and 0.146 eV. The vertical dashed lines mark the corresponding cooling threshold lengths $L_0 = (786, 1573, 3145)\lambda_p$ from Eq. (8b) of the main text. Beyond $L_0$, all lossless curves collapse onto the $\Delta E_f \propto 1/L_{int}$ dependence of Eq. (8a) of the main text. All curves are calculated with a photonic bandwidth of 1 eV (FWHM, raised-cosine edges, Sec. S4.5). For $\Delta E_i = 0.146$ eV, the curves labeled 5, 15 and 50 dB/cm show the saturation caused by photonic propagation loss, with the corresponding loss-limited floors $\Delta E_{f,sat} \approx 4.75$, 14.2 and 47.5 meV of Eq. (10) of the main text and Eq. (S4.6). Fig. 3a of the main text shows that in the lossless case, beyond $L_0$ the final spread depends only on $L_{int}$. The different cooling ratios show that each ratio drops below 1 near its own $L_0$, which scales as $1/\Delta E_i$, so a narrower initial spread needs a longer interaction length. With photonic loss, the ratio saturates at $\Delta E_{f,sat}/\Delta E_i$, which sets the largest cooling factor for a given loss. **(b)** Final electron energy spread $\Delta E_f$ (FWHM) as a function of $L_{int}/\lambda_p$ for $\Delta E_i = 0.584$ eV, $v_e = v_{p,g} = 0.1c$, $\lambda_p = 500$ nm and no propagation loss, for photonic bandwidths (FWHM, raised-cosine edges) of 0.3 eV, 0.6 eV, 1 eV (used in the main text) and 1.5 eV. The dotted black line shows the unbounded frequency band result of Eq. (8a) of the main text. The vertical dashed line marks $L_0 = 786\lambda_p$ from Eq. (8b) of the main text, and the horizontal dotted line marks $\Delta E_i$. Inset: enlarged view around $L_0$.

### S4.5 Finite Photonic Bandwidth

The numerical results of Figs. 2b, 2c and 3 and of Table 1 of the main text restrict the emitted photon frequency to a fixed photonic band centered on $\omega_{\mathrm{p}}$, within which the photonic dispersion is expanded to second order (Sec. S2.2). The band represents the frequency range over which the dispersion-engineered mode exists with the required group velocity and group-velocity dispersion, and it does not depend on $\Delta E_{\mathrm{i}}$. It is modeled by multiplying $\left|g_{\mathrm{eff},\omega}\right|^2$ by a raised-cosine profile of the photon energy, which equals 1 at $\omega_{\mathrm{p}}$, 1/2 at $\hbar\delta\omega = \pm 0.5$ eV and falls smoothly to zero at $\hbar\delta\omega = \pm 1$ eV. The photonic bandwidth quoted in the main text is the FWHM of this profile, 1 eV. No photon is emitted outside the band, so the band limits the photon energies that can carry away the electron energy spread.

The final spread is set by the phase-matching window, which narrows as $1/L_{\mathrm{int}}$ [Eq. (8a) of the main text]. At short interaction lengths, this window is wider than the band, and the band limits the final spread, so that the curves flatten instead of following Eq. (8a) of the main text. The numerical crossing $\Delta E_{\mathrm{f}} = \Delta E_{\mathrm{i}}$ therefore approaches the threshold length $L_0$ of Eq. (8b) of the main text only for a sufficiently wide band. Fig. S1b shows this for $\Delta E_{\mathrm{i}} = 0.584$ eV ($L_0 = 786\lambda_{\mathrm{p}}$). For a photonic bandwidth of 0.3 eV, narrower than $\Delta E_{\mathrm{i}}$, the band acts as an energy filter and the final spread stays below $\Delta E_{\mathrm{i}}$ at all lengths shown. For 0.6, 1 and 1.5 eV, the crossing lies at $556\lambda_{\mathrm{p}}$, $706\lambda_{\mathrm{p}}$ and $779\lambda_{\mathrm{p}}$, i.e. at $0.71L_0$, $0.90L_0$ and $0.99L_0$, respectively. At long interaction lengths, all bandwidths give the same $1/L_{\mathrm{int}}$ dependence, and at $L_{\mathrm{int}} = 2 \cdot 10^4\lambda_{\mathrm{p}}$ the final spreads agree to within 1%.

Figs. 2b, 2c and 3 of the main text and Fig. S1a use the photonic bandwidth of 1 eV. For the three initial spreads of Fig. S1a and Fig. 3a of the main text, $\Delta E_{\mathrm{i}} = 0.584$, 0.292 and 0.146 eV, the numerical crossing $\Delta E_{\mathrm{f}} = \Delta E_{\mathrm{i}}$ lies at 706, 1541 and $3131\lambda_{\mathrm{p}}$, compared with $L_0 = 786$, 1573 and $3145\lambda_{\mathrm{p}}$, i.e. 10%, 2% and 0.5% below $L_0$. The $\Delta E_{\mathrm{f}} = \Delta E_{\mathrm{i}}$ contour of Fig. 3b of the main text lies at $706\lambda_{\mathrm{p}}$ for $v_{\mathrm{e}} = v_{\mathrm{p,g}}$. The loss curves of Fig. 3a of the main text are unchanged by the band at long interaction lengths.

In Figs. 2b and 2c of the main text ($L_{\mathrm{int}} = 1$ cm), the band does not change the final spread at group-velocity matching: $\Delta E_{\mathrm{f}}/\Delta E_{\mathrm{i}} = 0.040$ at $v_{\mathrm{e}} = v_{\mathrm{p,g}}$ for photonic bandwidths between 0.6 and 2 eV. For $v_{\mathrm{e}} > v_{\mathrm{p,g}}$, the emitted photon spectrum is wider than the band, which truncates it, so that the height of the heating maximum depends on the bandwidth: $\Delta E_{\mathrm{f}}/\Delta E_{\mathrm{i}}$ reaches 1.4, 1.9, 2.5 and 2.9 for bandwidths of 0.6, 1, 1.5 and 2 eV. For 1 eV, the maximum lies at $v_{\mathrm{e}} = 0.10007c$, above the resonance $v_{\mathrm{e,f}} = v_{\mathrm{p,g}}$ of Eq. (6) of the main text ($v_{\mathrm{e}} = 0.10005c$), and the cooling–heating boundary $\Delta E_{\mathrm{f}} = \Delta E_{\mathrm{i}}$ lies at $v_{\mathrm{e}} = 0.10003c$, compared with $0.10002c$ from Eq. (7) of the main text. Fig. 2a of the main text is calculated without a band limit. The final spreads of Table 1 of the main text are calculated with the same 1 eV band, and with a FWHM of $2\Delta E_{\mathrm{i}} = 2$ MeV in the ultra-relativistic regime. Without a band limit, they change by less than 0.5%.

## S5 Relativistic Effects

### S5.1 Relativistic Electron Dispersion Relation

The relativistic energy-momentum relation is:

$$(pc)^2 = (E + mc^2)^2 - (mc^2)^2 = E^2 + 2Emc^2, \tag{S5.1}$$

where $p$ is the electron momentum, $E$ its kinetic energy and $m$ its rest mass. With $\gamma = 1 + E/mc^2$, the wavenumber and its first two derivatives are:

$$k(E) = \frac{p}{\hbar} = \frac{1}{\hbar c}\sqrt{E^2 + 2Emc^2} = \frac{mc}{\hbar}\sqrt{\gamma^2 - 1},$$
$$\frac{\partial k}{\partial E} = \frac{\gamma}{\hbar c\sqrt{\gamma^2 - 1}}, \quad \frac{\partial^2 k}{\partial E^2} = -\frac{1}{\hbar m c^3(\gamma^2 - 1)^{3/2}}. \tag{S5.2}$$

The quadratic term in $\delta E_{\mathrm{f}}$ also remains negligible in the relativistic regime, as in the non-relativistic case of Eq. (S2.5). For $\hbar\omega_{\mathrm{p}} \ll E_0$, its ratio to the linear term is $3\gamma|\delta E_{\mathrm{f}}|/2(\gamma + 1)E_0$, which reduces to the weak-recoil limit of Eq. (S2.5) for $\gamma \to 1$.

## S5.2 Relativistic Cooling Ratio and Saturation Floor

The cooling analysis of Sec. 2 of the main text carries over to relativistic electrons, with one change in the interaction. In the paraxial, spin-independent limit, minimal coupling gives $H_{\mathrm{int}} = e\mathbf{A} \cdot \hat{\mathbf{v}}$, where $\hat{\mathbf{v}}$ is the relativistic velocity operator, so that $\hbar k(E)/m$ in Sec. S1 is replaced by the relativistic velocity $v(E) = \partial E/\partial(\hbar k)$. Since Eqs. (S1.19), (S1.23) and (S1.26) are written in terms of $v(E)$, they hold unchanged with this velocity, and otherwise only the electron dispersion $k(E)$ changes. Eqs. (2) and (3) of the main text are written with $k(E)$ in place of $v(E)$. For relativistic electrons, the two forms of the prefactor of Eq. (3) differ by the factor $1 + \hbar\omega/\gamma mc^2$, which is $5 \cdot 10^{-6}$ at $v_{\mathrm{e}} = 0.1c$ and $3 \cdot 10^{-6}$ at 200 keV for $\lambda_{\mathrm{p}} = 500\, nm$, and about 1% in the 1 GeV regime of Table 1 of the main text. Across the emitted band it varies by about $\Delta E_{\mathrm{i}}/\gamma mc^2$, so it rescales the emission probability and does not change $\Delta E_{\mathrm{f}}$. With $\gamma = 1 + E_0/mc^2 = (1 - v_{\mathrm{e}}^2/c^2)^{-1/2}$, the electron velocity after emitting a photon of energy $\hbar\omega_{\mathrm{p}}$ is:

$$v_{\mathrm{e,f}} = c\sqrt{1 - \left(\gamma - \frac{\hbar\omega_{\mathrm{p}}}{mc^2}\right)^{-2}} = c\sqrt{1 - \left(\gamma - \frac{2\pi\hbar}{mc\lambda_{\mathrm{p}}}\right)^{-2}}. \tag{S5.3}$$

For $\hbar\omega_{\mathrm{p}} < \left(\gamma_{\mathrm{p,g}} - 1\right)mc^2$ with $\gamma_{\mathrm{p,g}} = \left(1 - v_{\mathrm{p,g}}^2/c^2\right)^{-1/2}$, for which group-velocity matching is possible, the cooling condition $v_{\mathrm{e}}^{-1} + v_{\mathrm{e,f}}^{-1} > 2v_{\mathrm{p,g}}^{-1}$ (Eq. (7) of the main text) becomes $v_{\mathrm{e}} < v_{\mathrm{p,g}}\left(1 + \frac{\pi\hbar c}{m\lambda_{\mathrm{p}}\gamma_{\mathrm{p,g}}^3 v_{\mathrm{p,g}}^2} + \cdots\right)$, which reduces to the non-relativistic expression $v_{\mathrm{e}} < v_{\mathrm{p,g}}\left(1 + \frac{\pi\hbar c}{m\lambda_{\mathrm{p}} v_{\mathrm{p,g}}^2} + \frac{\pi^2\hbar^2 c^2}{m^2\lambda_{\mathrm{p}}^2 v_{\mathrm{p,g}}^4} + \cdots\right)$ for $\gamma_{\mathrm{p,g}} \to 1$. At group-velocity matching, $v_{\mathrm{e}} = v_{\mathrm{p,g}}$, the group velocity difference term is $\Delta v_{\mathrm{i-f}}^{-1} = \frac{1}{c\sqrt{1-\left(\gamma-\frac{2\pi\hbar}{mc\lambda_{\mathrm{p}}}\right)^{-2}}} - v_{\mathrm{e}}^{-1}$, which reduces to the non-relativistic result for $\gamma \to 1$. In the weak-recoil limit $\hbar\omega_{\mathrm{p}} \ll E_0$, Eq. (S2.2) and the relativistic dispersion $\partial^2 k/\partial E^2 = -1/(\hbar\gamma^3 m v_{\mathrm{e}}^3)$ give:

$$\Delta v_{\mathrm{i-f}}^{-1} \approx \hbar^2\omega_{\mathrm{p}} \left.\frac{\partial^2 k}{\partial E^2}\right|_{E_0} = -\frac{\hbar\omega_{\mathrm{p}}}{\gamma(\gamma + 1)v_{\mathrm{e}}E_0} = -\frac{2\pi c\hbar}{\gamma(\gamma + 1)\lambda_{\mathrm{p}}v_{\mathrm{e}}E_0}, \tag{S5.4}$$

which reduces to $-\hbar\omega_{\mathrm{p}}/(2v_{\mathrm{e}}E_0)$ for $\gamma \to 1$. Inserting Eq. (S5.4) into Eq. (8a) of the main text and Eq. (S4.5) gives the relativistic cooling ratio and saturation floor, both larger than their non-relativistic values by the factor $\gamma(\gamma + 1)/2$:

$$\frac{\Delta E_{\mathrm{f}}}{\Delta E_{\mathrm{i}}} \simeq \frac{\mu\gamma(\gamma + 1)\lambda_{\mathrm{p}}v_{\mathrm{e}}E_0}{\pi c L_{\mathrm{int}}\Delta E_{\mathrm{i}}} = \frac{L_0}{L_{\mathrm{int}}}, \quad \frac{L_0}{\lambda_{\mathrm{p}}} \simeq \frac{\mu\gamma(\gamma + 1)v_{\mathrm{e}}E_0}{\pi c\Delta E_{\mathrm{i}}}, \tag{S5.5}$$

$$\Delta E_{\mathrm{f,sat}} \simeq \frac{\gamma(\gamma + 1)}{2}\frac{2v_{\mathrm{e}}E_0\alpha}{\omega_{\mathrm{p}}v_{\mathrm{p,g}}}. \tag{S5.6}$$

As in the non-relativistic case, the floor is independent of $\Delta E_{\mathrm{i}}$ and $L_{\mathrm{int}}$ and depends only on the spatial attenuation coefficient $\alpha/v_{\mathrm{p,g}}$.

### S5.3 Relativistic Dispersion-Engineering Conditions

The dispersion-engineering conditions of Sec. S2.2 are written in terms of $k(E)$, so they hold for the relativistic dispersion of Eq. (S5.2). Group-velocity matching, $v_{\mathrm{p,g}} = v_{\mathrm{e}}$, is unchanged, with $v_{\mathrm{e}} = c(1-\gamma^{-2})^{1/2}$. With $\gamma^2 - 1 = \gamma^2 v_{\mathrm{e}}^2/c^2$, Eq. (S5.2) gives the required group-velocity dispersion of the photonic mode:

$$\left.\frac{\partial^2 q}{\partial \omega^2}\right|_{\omega_{\mathrm{p}}} = \hbar^2 \left.\frac{\partial^2 k}{\partial E^2}\right|_{\mathrm{E_0}} = -\frac{\hbar}{\gamma^3 m v_{\mathrm{e}}^3}, \tag{S5.7}$$

which is anomalous at all energies. Its magnitude is smaller than the non-relativistic value $\hbar/mv_{\mathrm{e}}^3$ by the factor $\gamma^3$, because the electron velocity changes less with energy as $v_{\mathrm{e}}$ approaches $c$. It is 4.2 ps²/km at $v_{\mathrm{e}} = 0.1c$, 0.022 ps²/km at 80 keV and $6 \cdot 10^{-13}$ ps²/km at 1 GeV, so the requirement on the photonic dispersion relaxes at high energies. Phase matching requires $q(\omega_{\mathrm{p}}) = k(E_0) - k(E_0 - \hbar\omega_{\mathrm{p}})$. Expanding the last term to second order in $\hbar\omega_{\mathrm{p}}$, with $\partial k/\partial E = 1/\hbar v_{\mathrm{e}}$ and Eq. (S5.7), gives:

$$q(\omega_{\mathrm{p}}) \approx \frac{\omega_{\mathrm{p}}}{v_{\mathrm{e}}} + \frac{\hbar\omega_{\mathrm{p}}^2}{2\gamma^3 m v_{\mathrm{e}}^3} = \frac{\omega_{\mathrm{p}}}{v_{\mathrm{e}}}\left[1 + \frac{\hbar\omega_{\mathrm{p}}}{2\gamma(\gamma+1)E_0}\right], \tag{S5.8}$$

where $\gamma^3 m v_{\mathrm{e}}^2 = \gamma(\gamma+1)E_0$ was used. The neglected third-order term is smaller than the second-order term by the factor $\gamma\hbar\omega_{\mathrm{p}}/(\gamma+1)E_0$, which is $5 \cdot 10^{-4}$ at $v_{\mathrm{e}} = 0.1c$ and $\lambda_{\mathrm{p}} = 500$ nm and smaller at higher energies. The phase velocity of the mode, $\omega_{\mathrm{p}}/q(\omega_{\mathrm{p}})$, therefore lies below $v_{\mathrm{e}}$ by the relative amount $\hbar\omega_{\mathrm{p}}/2\gamma(\gamma+1)E_0$, which by Eq. (S5.4) places it midway between $v_{\mathrm{e}}$ and $v_{\mathrm{e,f}}$. The truncation condition of Eq. (S2.6) is evaluated with $\partial^3 k/\partial E^3 = 3/\hbar\gamma^4 m^2 v_{\mathrm{e}}^5$, so the electron contribution to the residual cubic phase decreases as $\gamma^{-4}$.

## S6 Comparison with Conventional Monochromation

The relevant comparison is the useful current delivered at a specified energy width. Let $I_{\mathrm{cool}}$ denote the current of the accepted cooled population, including selection and transport efficiencies, and let $I_{\mathrm{mono}}$ denote the current transmitted by a conventional monochromator at comparable energy resolution. For equal incident current, acquisition time, and detection efficiency, a shot-noise-limited comparison with the ratio of the two SNRs is: $\frac{\mathrm{SNR_{cool}}}{\mathrm{SNR_{mono}}} = \sqrt{\frac{I_{\mathrm{cool}}}{I_{\mathrm{mono}}}}$. An advantage for the same energy width therefore requires $I_{\mathrm{cool}} > I_{\mathrm{mono}}$.

### S6.1 Monochromator: Flux and SNR

Here we derive this comparison for a Gaussian initial energy distribution. Consider a beam of total current $I_0$ with a Gaussian energy distribution of FWHM $\Delta E_{\mathrm{i}}$:

$$f(E) = \frac{2I_0}{\Delta E_{\mathrm{i}}}\sqrt{\frac{\ln 2}{\pi}}\exp\left(-4\ln 2\frac{(E-E_0)^2}{\Delta E_{\mathrm{i}}^2}\right). \tag{S6.1}$$

The cooling tradeoff of conventional monochromators: a monochromator uses an energy-dispersive element followed by a slit that transmits only the electrons within a window $\Delta E_{\mathrm{f}} \ll \Delta E_{\mathrm{i}}$ around $E_0$:

$$I_{\mathrm{mono}} = f(E_0)\Delta E_{\mathrm{f}} = 2\sqrt{\frac{\ln 2}{\pi}}\frac{I_0 \Delta E_{\mathrm{f}}}{\Delta E_{\mathrm{i}}} \approx \frac{I_0 \Delta E_{\mathrm{f}}}{1.06\Delta E_{\mathrm{i}}}. \tag{S6.2}$$

For shot-noise-limited detection, $\mathrm{SNR} \propto \sqrt{I}$, so the monochromator SNR degrades as $\sqrt{\Delta E_{\mathrm{f}}}$ with finer resolution.

### S6.2 Nanophotonic Cooling: Flux and SNR

The final electron state after the interaction is split into two parts. Electrons that did not emit remain unchanged, and electrons that did emit are shifted down in energy by roughly $\hbar\omega_{\mathrm{p}}$. Each electron spontaneously emits into the vacuum photonic mode with probability $P_{\mathrm{em}}$ of Eq. (S1.24), where in the weak-coupling regime $|g_{\mathrm{Q}}|^2 \ll 1$, the single-photon emission probability scales as $P_{\mathrm{em}} \approx |g_{\mathrm{Q}}|^2$, and $P_{\mathrm{em}} \leq 1$ in general. The cooled sub-ensemble is post-selected from the zero-loss peak, so that:

$$I_{\mathrm{cool}} = P_{\mathrm{em}} I_0, \tag{S6.3}$$

independently of the final width $\Delta E_{\mathrm{f}}$. In contrast to conventional monochromators, this post-selection does not set the final energy width, which is set by the cooling process itself, removing the intrinsic resolution-flux tradeoff that exists in monochromators.

### S6.3 SNR Ratio and Crossover

The ratio of the two SNRs is:

$$\frac{\mathrm{SNR}_{\mathrm{cool}}}{\mathrm{SNR}_{\mathrm{mono}}} = \sqrt{\frac{I_{\mathrm{cool}}}{I_{\mathrm{mono}}}} = \sqrt{1.06 P_{\mathrm{em}} \frac{\Delta E_{\mathrm{i}}}{\Delta E_{\mathrm{f}}}}, \tag{S6.4}$$

so that cooling outperforms the monochromator when:

$$\frac{\Delta E_{\mathrm{f}}}{\Delta E_{\mathrm{i}}} < 1.06 P_{\mathrm{em}}, \tag{S6.5}$$

For example, with $P_{\mathrm{em}} \sim 10^{-2}$ and $\Delta E_{\mathrm{i}} \sim 1$ eV, cooling is advantageous for final widths below $\approx 10$ meV (which is close to the current monochromator state of the art). Of course, the perturbative regime persists even up to $P_{\mathrm{em}} \sim 10^{-1}$, so an advantage for cooling may be possible even for final widths below $\approx 100$ meV. Indeed, cooling may persist beyond the perturbative regime, which would extend even further the advantage gained from cooling, but that is beyond the scope of the current manuscript.

## S7 Potential Cooling Performance

### S7.1 Cooling Performance Across Energy Regimes

Reducing the energy spread is beneficial for many different applications of electron microscopy, spectroscopy, and beam physics, spanning a large range of electron energies. We now examine each application by its typical electron energy range and characterize these representative energy regimes by their typical initial energy spreads and determine their required properties of the photonic structure, concluding the achievable cooling ratio for each regime. In general, following Eq. ($8a$) of the main text, the cooling ratio at group-velocity matching is better for slower electrons, but they also require slow light. The cooling ratio is also better for shorter wavelength $\lambda_p$. We consider practical achievable limits on the interaction lengths, limiting the cooling ratio for each regime, and its dispersion requirements in Sec. S7.2.

Low-energy electrons ($\sim 20 - 300\ eV$) and SEM-scale energies ($\sim 1 - 30\ keV$) emitted from conventional thermionic or field-emission guns exhibit FWHM energy spreads of $0.6 - 0.8\ eV$ for a Schottky emission (SE)-type emitter and $0.3\ eV$ for a cold field emission emitter [13,14]. These spreads are relatively large compared with the electron beam energy at these low-energy regimes and represent a known barrier to aberration correction [15], severely limiting resolution and contrast. At these velocities, coupling to visible/near-IR wavelengths nanophotonic structures, such as $\lambda_p = 500\ nm$, is feasible via Smith-Purcell emission from nanoscale gratings [16] and via inverse Smith–Purcell from dielectric grating interactions [17]. Group-velocity-matching at these low energies, however, requires slow-light modes that remain nearly constant over the emitted bandwidth.

In scanning electron microscopy (SEM) operating at low primary energies (1-5 keV), the electron energy spread represents a substantial fraction of the total beam energy. This energy uncertainty fundamentally limits aberration correction. Conventional monochromators achieve this by energy filtering, at the expense of electron current. For example, [18] experimentally demonstrated a reduction in energy spread from 770 to 67 meV (FWHM) using an offset cylindrical-lens monochromator, with only 5.7% electron transmission. For our representative 2.5-keV SEM parameters (Table 1 of the main text), an initial energy spread of 0.6 eV, a photon wavelength of 500 nm, and a 1-cm interaction length yield a predicted final FWHM of approximately $\Delta E_f = 22$ meV, about 3× narrower. Unlike conventional energy filtering, our scheme determines the final energy width through the electron-photon interaction rather than the width of a selection slit. The useful electron current is instead determined by the spontaneous-emission probability and the collection efficiency of the cooled population. Consequently, an effective cooled-electron yield exceeding 5.7% would provide greater useful current than the reported conventional monochromator, assuming comparable incident beam currents and collection conditions. These results suggest a route toward simultaneously improving energy resolution and electron throughput in SEM, provided that the required dispersion engineering, interaction length, and emission efficiency can be realized experimentally.

In electron energy-loss spectroscopy (EELS), where the electron energies are typically 20–100 keV, accessing weak excitonic or vibrational resonances requires that the feature linewidth substantially exceeds the source energy spread [19,20]. For example, low-energy acoustic phonons in transition metal dichalcogenides (e.g., $MoS_2$) exhibit linewidths approximately $0.5 - 1$ meV, which are considerably narrower than the typical $\sim 0.6$ eV energy spread from conventional Schottky-emission electron guns. State-of-the-art monochromators can now reach sub 10 meV at 60-100 keV, and it is improved to sub 5 meV at lower energies of 20-30 keV [21–23]. Reaching sub 5 meV for 20 keV electrons with our cooling scheme requires a very short photon wavelength in the EUV range and an interaction length of at least 20 cm.

TEM energies ($\sim 30 - 300\ keV$) are the workhorse regime for atomic-resolution imaging and spectroscopy, where the energy spread affects both chromatic aberration and spectral resolution. A major advantage of this regime is that strong, coherent electron-light coupling is already extensively demonstrated, especially through photon-induced near-field electron microscopy (PINEM) and related cavity or waveguide platforms at telecom wavelengths [24,25]. At intermediate relativistic energies ($\sim 200\ keV$), modern photonic techniques have demonstrated coupling to optical fields with sub-eV energy spread. Experiments have directly demonstrated extended resonant phase matching between propagating light fields and free-electron wavefunctions [26].

In the ultra-relativistic accelerator regime ($\sim 1\ GeV$), relevant to high-brightness beams for free-electron lasers (FELs) and compact advanced accelerators, the energy spread arises from RF curvature, residual energy chirp, timing jitter, and collective effects such as space charge and coherent synchrotron radiation. Energy spread is strongly source-dependent: plasma-based GeV beams often begin at the percent level unless actively conditioned, and recent work explicitly targets sub-percent operation, e.g., $0.17\%$ at $1\ GeV$ [27], or $\Delta E_\mathrm{i} \sim 1\ MeV$ in compact linacs or FELs [28] via chirp compensation. Since cooling requires $\hbar\omega_\mathrm{p} > \Delta E_\mathrm{i}$ (Sec. 2 of the main text), the MeV-scale energy spread of these beams calls for far higher photon energies than in the microscopy regimes. Intracavity Compton backscattering in storage-ring free-electron lasers has been used to generate high-brightness Gamma-rays ($\sim 5 - 35\ MeV$), demonstrating access to photon energies suitable for cooling in this regime [29,30]. Since these photon energies remain small compared with the electron energy, $\hbar\omega_\mathrm{p}/E_0 \approx 1\%$, the emission lies within the relativistic weak-recoil limit of Sec. S5.2, and Eq. (S5.5) gives a threshold length of only $L_0 \approx 0.3$ mm. This estimate, however, assumes a structure that phase- and group-velocity matches the electron at $\gamma$-ray frequencies, which has no current photonic realization: in vacuum, the group-velocity mismatch $v_\mathrm{p,g} - v_\mathrm{e} \approx c/(2\gamma^2)$ exceeds the photon recoil by a factor of $\sim 40$ for these parameters, so that Eq. (6) of the main text gives a cooling ratio of $\approx 0.98$, i.e., essentially no cooling. Nevertheless, the required slowing is very small, $v_\mathrm{p,g} \approx c\left(1 - \frac{1}{2\gamma^2}\right)$, i.e., a fractional reduction of only $\sim 10^{-7}$, and control over the group velocity of high-energy photons is emerging: the group velocity of 14.4 keV X-rays has already been reduced by four orders of magnitude using Mössbauer nuclei embedded in a thin-film cavity [31]. Extending such dispersion engineering to broadband MeV photons could open the ultra-relativistic regime to passive nanophotonic cooling in the future.

## S7.2 Dispersion requirements and candidate implementations

Implementation of the proposed cooling mechanism must satisfy phase matching, group-velocity matching, and recoil-compensating dispersion over the relevant emission bandwidth. For strong energy compression, the emitted photon spectrum carries approximately the initial electron energy spread, so its bandwidth is of order $\hbar\delta\omega \sim \Delta E_\mathrm{i}$. For example, $\Delta E_\mathrm{i} = 0.5$ eV corresponds to a fractional frequency bandwidth of approximately 14% at 350 nm. The photonic dispersion must therefore be assessed across the emission band, including higher-order terms, rather than only at its central frequency. Experiments on slow-light photonic-crystal waveguides demonstrate that reduced group velocity can accompany substantial higher-order dispersion [32].

For relativistic free electrons, the required second-order dispersion of $\left.\frac{\partial^2 q}{\partial \omega^2}\right|_{\omega_\mathrm{p}} = \hbar^2 \left.\frac{\partial^2 k}{\partial E^2}\right|_{E_0}$, as discussed in Sec. S2.2, is $\left.\frac{\partial^2 q}{\partial \omega^2}\right|_{\omega_\mathrm{p}} = \hbar^2 \left.\frac{\partial^2 k}{\partial E^2}\right|_{E_0} = -\hbar/(\gamma^3 m v_\mathrm{e}^3)$ (Sec. S5.3), which is negative and small in magnitude. The design objective is therefore a mode with the prescribed group velocity and weak anomalous curvature across the relevant bandwidth. Shorter wavelengths reduce the required fractional bandwidth and improve the finite-length cooling scale. The condition $\hbar\omega_\mathrm{p} > \Delta E_\mathrm{i}$, adopted here to separate the emitting and non-emitting populations, additionally requires $\lambda_\mathrm{p} < hc/\Delta E_\mathrm{i}$ .

### *S7.2.1 SEM-energy electrons*

Silicon nitride (SiN) on insulator is a candidate platform because it combines low propagation loss with established fabrication methods and demonstrated coupling to free electrons. Liu et al. reported SiN microresonators with losses corresponding to approximately 1 dB/m, together with meter-long waveguides exhibiting a few dB/m of loss [33]. Henke et al. demonstrated coherent electron phase modulation using an optically driven SiN microresonator [34]. These results establish relevant fabrication and coupling capabilities, although neither establishes the simultaneous dispersion and bandwidth requirements of passive cooling.

Fused silica ($SiO_2$) provides another candidate platform. Ref. [35] demonstrated synchronous interaction of 28 keV electrons with an optically driven fused-silica grating. This supports the feasibility of electron access and phase synchronism near dielectric gratings. Dispersion engineering has also been demonstrated in silica ridge waveguides, including anomalous dispersion and generation of ultraviolet dispersive waves [36]. Ultraviolet operation is attractive because increasing photon energy reduces both the finite-length width and the loss floor at fixed attenuation per unit length. Silica-based ultraviolet guidance is experimentally established [37]. As an illustrative assumption, taking $\lambda_\mathrm{p} = 350$ nm and a propagation loss of 3 dB/cm gives saturation widths of approximately 0.48 and 5.4 meV at electron energies of **1 and 5 keV**, respectively. The corresponding lossless lengths required for a 10 meV width are approximately 3.9 mm and 4.4 cm, respectively. These estimates identify a favorable energy–wavelength regime, but do not constitute predictions for an existing device. Shorter wavelengths also reduce the evanescent decay length of the synchronous field, making electron proximity and alignment more demanding.

Using Eq. (10) of the main text, a propagation loss of 1 dB/cm gives an estimated saturation width of approximately 5.2 meV for 5 keV electrons at 1 $\mu$m. Reducing the loss to 0.1 dB/cm (far above the lowest loss of silica fibers [38]) lowers this asymptotic estimate to approximately 0.52 meV. However, these values describe the long-interaction limit. Eq. (8a) of the main text predicts a lossless interaction length of approximately 12.5 cm to reach 10 meV under the same conditions. Low optical loss is therefore necessary but insufficient: maintaining electron–mode overlap and the prescribed dispersion over the required length is an additional constraint.

### *S7.2.2 TEM-energy electrons*

A near-term demonstration at TEM energies does not require reaching the threshold length. For 80 keV electrons and a periodic fused-silica hole-array structure, lowering the photon wavelength to UV photons with $\lambda_\mathrm{p} \approx 200$ nm, still within the transparency window of UV-grade fused silica [39,40], reduces $L_0$ to about 3 cm, compared with 7 cm at 500 nm (Table 1 of the main text). Since $1/\beta \approx 1.99$ exceeds the refractive index of silica, phase matching must rely on a spatial harmonic of the periodic structure [41], while the required group-velocity dispersion, about −0.02 ps²/km, is essentially zero. Instead of the cooled width, one can measure the slope $\delta E_\mathrm{f}/\delta E_\mathrm{i}$ of Eq. (6) of the main text by stepping the accelerating voltage by a few tenths of a volt ($\delta \approx \pm 0.1$ eV, about 1 ppm of 80 kV) with the spectrometer settings fixed. The zero-loss peak then follows the beam with unit slope, whereas the cooled peak shifts only by $(\delta E_\mathrm{f}/\delta E_\mathrm{i})\delta$. Peak positions can be determined far more precisely than peak widths, so this works even when the cooled peak is still broader than $\Delta E_\mathrm{i}$. A slope of $|\delta E_\mathrm{f}/\delta E_\mathrm{i}| \leq 0.5$ is obtained when $v_\mathrm{p,g}$ lies between $v_\mathrm{e} - (v_\mathrm{e} - v_\mathrm{e,f})/3$ and $v_\mathrm{e} + (v_\mathrm{e} - v_\mathrm{e,f})$, where $(v_\mathrm{e} - v_\mathrm{e,f})/v_\mathrm{e} \approx \hbar\omega_\mathrm{p}/\gamma(\gamma+1)E_0 \approx 3 \cdot 10^{-5}$ is the relative velocity change of the electron upon photon emission. At the upper bound, the slope is +0.5 and the threshold length is halved. At the lower bound, the slope is −0.5, and the window is asymmetric because the heating resonance, $v_\mathrm{p,g} = v_\mathrm{e,f}$, lies on the slow side. The minimum interaction length is set by the requirement that the emitted photons remain inside the engineered band. These photons are spread over $(1 - \delta E_\mathrm{f}/\delta E_\mathrm{i})(\Delta E_\mathrm{i} + 2|\delta|)$ plus the phase-matching window $(1 - \delta E_\mathrm{f}/\delta E_\mathrm{i})\Delta E_\mathrm{i} L_0/L_\mathrm{int}$, and truncation at the band edges would bias the measured slope toward unity. For a ±1 eV band around 6 eV, this gives $L_\mathrm{int} \approx 6$ mm at $\delta E_\mathrm{f}/\delta E_\mathrm{i} = 0.5$. This operating point is not optimal for cooling, since half of each electron's initial deviation is retained and the cooled width saturates at about $0.5\Delta E_\mathrm{i}$ for long interaction lengths, but it suffices to demonstrate the mechanism, as a slope of 0.5 is readily distinguished from unity. At exact matching, the same band requires $L_\mathrm{int} \approx 1.5$ cm. Over these lengths the beam must remain within the evanescent decay length $\beta\gamma\lambda_\mathrm{p}/2\pi \approx 20$ nm of the structure [42,43], beyond typical unguided aloof interaction lengths, so ponderomotive guiding [44] is likely required. Two practical points follow. First, a relative fabrication error in the group velocity shifts the operating energy by $\gamma(\gamma+1)E_0$ times that error, about 200 eV per $10^{-3}$, so a coarse voltage search must precede the fine scan. Second, spectra accumulated over many frames should not be aligned to the zero-loss peak, which would smear the cooled peak. Frame-by-frame correlation of the two peak positions under natural high-voltage fluctuations can instead replace the deliberate scan.

For relativistic electron energies of 100–300 keV, the weak-recoil phase-matching requirement is $n_{\mathrm{eff}} = \frac{c}{v_{\mathrm{e}}} \simeq 1.82 - 1.29$. These effective indices are accessible in dielectric photonic structures. Extended phase-matched electron–light interaction and coupling to integrated microresonators have been demonstrated [26,34], making this regime attractive for testing the interaction mechanism. Nevertheless, the required effective index must be obtained from the guided mode and matching it does not automatically establish the required group index and dispersion.

The smaller recoil-induced velocity change at these energies increases both the required interaction length and sensitivity to propagation loss. At 200 keV, a loss of 0.5 dB/cm gives estimated saturation widths of approximately 0.85 eV at 1 $\mu$m and 0.42 eV at 500 nm. Both exceed the lossless final spread of the 200 keV regime in Table 1 of the main text (41 meV), which therefore requires a propagation loss well below 0.05 dB/cm. At the propagation loss of 0.1 dB/km considered in Sec. 2 of the main text, Eq. (S5.6) gives a floor of 0.85 $\mu$eV at 200 keV and 500 nm, and the floor of about 1 neV quoted there holds for electron energies up to about 10 keV with $\lambda_{\mathrm{p}} \gtrsim 150$ nm (0.9 neV at 2.5 keV and 500 nm, 2 neV at 10 keV and 150 nm). Even without loss, reaching 100 meV at 1 $\mu$m requires an interaction length of approximately 4.1 m within the present model. Substantial TEM-energy cooling therefore requires significantly lower modal attenuation and a geometry supporting much longer effective interaction lengths.

### S7.3 Other charged-particle cooling

Our formalism may also motivate studies of other charged particles, but their achievable cooling depends on particle mass, velocity, radiative coupling, and beam geometry. The Muon Ionization Cooling Experiment (MICE) [45] demonstrated transverse cooling through energy loss in an absorber, a mechanism distinct from the mode-selective spontaneous emission considered here. Extending the present scheme to muons or other particles therefore requires a separate quantitative analysis.

## Notation Table

| Symbol | Definition | Defined in |
|---|---|---|
| $e, m, \hbar, \epsilon_0, c$ | Elementary charge, electron mass, reduced Planck constant, vacuum permittivity, speed of light | Sec. S1.1 |
| $E, E_0$ | Electron kinetic energy, central beam energy | Sec. S1.1, S1.6 |
| $H, H_{\mathrm{e}}, H_{\mathrm{ph}}, H_{\mathrm{int}}$ | Total, electron, photon and interaction Hamiltonians | Eqs. (S1.1)–(S1.5) |
| $\hat{a}_{\omega}, \mathbf{A}$ | Photon annihilation operator, quantized vector potential | Eqs. (S1.3), (S1.8) |
| $\psi_{\perp}, \tilde{u}_{\omega}^{(\mathrm{z})}$ | Normalized transverse electron wavefunction and transverse mode profile | Eqs. (S1.6), (S1.10) |
| $\mathbf{r}_{\perp}, \mathbf{r}_{\mathrm{T,e}}$ | Transverse coordinate, transverse position of the beam center | Sec. S1.1, S3.2 |
| $E_{\mathrm{i}}, E_{\mathrm{f}}$ | Initial and final electron energy | Sec. S1.4 |
| $\Delta E_{\mathrm{i}}, \Delta E_{\mathrm{f}}$ | Initial and final electron energy spread (FWHM) | Eq. (S1.25), Sec. S2.2 |

| Symbol | Definition | Defined in |
|---|---|---|
| $\delta E_{\mathrm{i}}, \delta E_{\mathrm{f}}, \delta\omega$ | Deviations from the phase-matched point | Sec. S2.1 |
| $k(E), v(E), v_{\mathrm{e}}$ | Electron longitudinal wavenumber and velocity, central electron velocity | Sec. S1.1, Eq. (S1.7), Sec. S2.2 |
| $v_{\mathrm{e,f}}$ | Electron velocity after emitting a photon of energy $\hbar\omega_{\mathrm{p}}$ | Sec. S2.2 |
| $\omega, \omega_{\mathrm{p}}, \lambda_{\mathrm{p}}$ | Photon frequency, phase-matched frequency and wavelength | Sec. S1.2, S1.6, S2.2 |
| $q(\omega), v_{\mathrm{p,g}}$ | Dispersion relation and group velocity of the photonic mode | Sec. S1.2 |
| $\partial^2 q/\partial\omega^2$ | Group-velocity dispersion of the photonic mode | Sec. S2.2 |
| $\Delta k$ | Phase mismatch | Eq. (S1.18) |
| $\Delta v_{\mathrm{GVM}}^{-1}$ | Group-velocity mismatch, $v_{\mathrm{e}}^{-1} - v_{\mathrm{p,g}}^{-1}$ | Eq. (S2.2) |
| $\Delta v_{\mathrm{i-f}}^{-1}$ | Recoil velocity mismatch, $v_{\mathrm{e}}^{-1} - v_{\mathrm{e,f}}^{-1}$ | Eq. (S2.2) |
| $L_{\mathrm{int}}$ | Interaction length | Sec. S1.4 |
| $L_0$ | Threshold interaction length for cooling | Eq. (8a) of the main text |
| $\mu$ | FWHM factor of the sinc$^2$ kernel | Sec. S2.2 |
| $g_{\mathrm{eff},\omega}$ | Transverse electron–photon overlap coefficient (inverse length) | Eq. (S1.16) |
| $M_{E_{\mathrm{i}}\to E_{\mathrm{f}},\omega}$ | First-order transition matrix element | Eq. (S1.19) |
| $\rho_{\mathrm{i}}, \rho_{\mathrm{f}}$ | Initial and final electron density matrix | Sec. S1.5, Eq. (S1.23) |
| $\left\lvert g_{\mathrm{Q}}\right\rvert^2$ | Electron–photon coupling constant (emission probability) | Eq. (S1.24) |
| $P_{\mathrm{em}}$ | Single-photon emission probability per electron, bounded by 1 | Eq. (S1.24) |
| $\beta$ | Electron velocity in units of the speed of light | Sec. S7.2.2 |
| $\alpha$ | Photonic loss rate (mode linewidth) | Sec. S3.5, S4.1 |
| $\Delta E_{\mathrm{f,sat}}$ | Loss-limited saturation energy spread | Sec. S4.2, S4.4 |
| $\mathcal{L}(x)$ | Normalized Lorentzian line shape of the photonic mode | Eq. (S4.1) |
| $L_{\mathrm{loss}}$ | Loss-limited effective interaction length | Sec. S4.4, Sec. 2 of the main text |
| $\gamma$ | Lorentz factor | Sec. S5.1 |
| $\gamma_{\mathrm{p,g}}$ | Lorentz factor of an electron moving at the group velocity | Sec. S5.2 |
| $\mathbf{G}(\mathbf{r}, \mathbf{r}', \omega), \mathbf{f}(\mathbf{r}, \omega)$ | Dyadic Green's function, MQED bosonic noise-field operator | Sec. S3.1 |

| Symbol | Definition | Defined in |
|---|---|---|
| $\varepsilon(\mathbf{r},\omega)$ | Complex permittivity of the photonic structure | Sec. S3.1 |
| $\mathbf{u}_n(\mathbf{r}), \omega_n, \alpha_n$ | Cavity modes, their frequencies and decay rates | Sec. S3.5 |
| $I_0, I_{\mathrm{cool}}, I_{\mathrm{mono}}$ | Incident, cooled and monochromator-transmitted currents | Sec. S6 |
| $n_{\mathrm{eff}}$ | Effective index required for phase matching | Sec. S7.2.2 |
| $\delta$ | Energy step of the accelerating-voltage scan | Sec. S7.2.2 |